# Theory of Faradaically Modulated Redox Active Electrodes for Electrochemically Mediated Selective Adsorption Processes

Fan He[1], Martin Z. Bazant[1,2], T. Alan Hatton[1]*

[1]Department of Chemical Engineering, Massachusetts Institute of Technology, Cambridge, Massachusetts 02139, USA.

[2]Department of Mathematics, Massachusetts Institute of Technology, Cambridge, Massachusetts 02139, USA

*Corresponding author: tahatton@mit.edu

**Abstract**

Electrochemically mediated selective adsorption is an emerging electrosorption technique that utilizes Faradaically enhanced redox active electrodes, which can adsorb ions not only electrostatically, but also electrochemically. The superb selectivity (>100) of this technique enables selective removal of toxic or high-value target ions under low energy consumption. Here, we develop a general theoretical framework to describe the competitive electrosorption phenomena involving multiple ions and surface-bound redox species. The model couples diffusion, convection and electromigration with competitive surface adsorption reaction kinetics, consistently derived from non-equilibrium thermodynamics. To optimize the selective removal of the target ions, design criteria were derived analytically from physically relevant dimensionless groups and time scales, where the propagation of the target anion's concentration front is the limiting step. Detailed computational studies are reported for three case studies that cover a wide range of inlet concentration ratios between the competing ions. And in all three cases, target anions in the electrosorption cell forms a self-sharpening reaction-diffusion wave front. Based on the model, a three-step stop-flow operation scheme with a pure stripping solution of target anions is proposed that optimizes the ion adsorption performance and increases the purity of the regeneration stream to almost 100%, which is beneficial for downstream processing.

**Introduction**

The treatment of natural water resources and process streams for the removal of trace compounds is of increasing importance to ensure safe drinking water, minimize discharge of pollutants with effluent process streams, and provide high-purity water for sensitive industrial operations [1–3]. In many instances the targeted compound will be in solution with a vast excess of other harmless salts, e.g., many micropollutants are found in water resources because they are often not responsive to traditional purification methods [4–6]. A number of existing water treatment processes either require substantial addition of chemicals during regeneration (e.g., adsorption or ion exchange), which generates a secondary waste stream; or are relatively non-specific for targeted species (e.g., desalination processes such as reverse osmosis, electrodialysis and multi-stage distillation). To overcome these limitations, there has been increasing interest in electrosorption processes in which high surface area electrodes are activated for adsorption of the targeted compounds under mild operating conditions (at room temperature, atmospheric pressure and around 1 V); during the regeneration step, the electrodes can then be deactivated on simple reduction of the cell voltage followed by release of the adsorbed compounds to a sweep stream thereby avoiding the traditional need for introduction of harsh eluent chemicals. One familiar example is capacitive deionization, for the desalination of brackish waters, which has undergone rapid development over the past two decades [7–13]. This technique, however, generally lacks specificity for a range of organic micropollutants of interest. For this reason, our group has been developing new approaches based on Electrochemically Mediated Selective Adsorption (EMSA) [14–18], an emerging technique in which reversible ion adsorption onto appropriately-selected redox active electrodes is mediated by Faradaic electron transfer reactions. This approach is related to, but is materially different from other examples of "pseudo-" [19–21] or "Faradaic" [22–26] capacitive deionization (CDI) approaches reported in the literature. Since the EMSA concept was first introduced as a general method of molecular separations from ionic solutions [18], it has been applied to selective removal of micropollutants such as organic acids, personal care products, pharmaceuticals [14,15,27], and heavy metal oxyanions [17,28] in wastewater remediation.

The operating principle of EMSA shares common features with other electrosorption methods, such as electrostatic (electric double-layer based) capacitive deionization, but has the major advantage that it can be tailored for strong electrochemical selectivity towards target ions for specific applications through appropriate selection of electrode materials and applied voltages [27]. Traditional electrostatic CDI is usually performed with porous carbon electrodes based on the non-Faradaic electrosorption of counter-ions in the electric double layers [13], and only weak, time-dependent ion selectivity is possible by exploiting nonlinearities in double layer charging dynamics [12,29]. Recent progress has been made in the functionalization of carbon electrodes to enable higher ion selectivity towards certain anions [30] and cations [31,32]. In this context, Faradaic reactions, if present at all, play only a secondary role, usually associated with undesirable side reactions and electrode degradation [24,25,33–35]. In contrast, EMSA leverages the unique electrochemical response of redox active materials, such as redox polymers [14–18,27,36], metal oxides [37,38], hexacyanoferrate (and its derivatives) [39–41] and other functionalized carbon composite materials [42] in order to achieve tunable, selective separations. In EMSA, ions are not only stored electrostatically by capacitive charging of the electric double layers, but also electrochemically by "pseudo-capacitive" charging of the electrodes through Faradaic reactions [22,27]. Strong selectivity can be achieved via either the specific binding of target ions or molecules to the active sites on the electrode surface [14,17,27] or the intercalation of target species into the crystal lattice of the electrode [36–41], both of which are mediated by Faradaic electrosorption reactions. In most cases of EMSA, capacitive double layer charging makes only a small contribution to the total ion removal, compared to the more selective ion removal by Faradaic reactions [17,27,43].

Although EMSA opens promising new directions for selective separations, most of the work to date has been on material development and characterizations in batch systems, and there has been little theoretical analysis on the overall process performance that could be used to design electrosorption systems to optimize selectivity towards target ions under flow conditions [22,27]. Moreover, in most capacitive deionization studies, the salt removal percentage was between 20~50% to avoid high energy consumption under low ionic concentrations [10,13,33], whereas for practical applications, complete depletion of the target ionic micropollutants is essential to achieve high selectivity towards the target anions as well as purity in the regenerated effluent or product streams. To achieve the desired separation performance, a fundamental understanding of the multi-ion transport phenomena in the electrosorption process is vital.

To achieve effective separation of ionic species in a continuous flow reactor, there are several crucial considerations that need to borne in mind. Beyond choosing electrode materials with high solute adsorption capacity and high affinity towards the target ions vs. the competing ions, proper design of the electrosorption cell is also essential to enable the complete depletion of the target ions, and should include considerations such as high surface to volume ratio and reduced dispersion effects, which is difficult to achieve in batch reactors. Moreover, the operational parameters need to be chosen carefully to ensure that the effective space velocity of the target ions is the slowest among all competing ions, which allows a clear separation of their concentration fronts. This is analogous to the design principle in packed bed adsorption [44,45], or displacement chromatographic columns [46,47]. During desorption, on the other hand, in addition to the regeneration concentration, the purity of the regenerated target ions is another important metric of practical interest, which can in principle be optimized by suitable choices for the stripping stream as well as the operating scheme during the washing and desorption steps. The ability to achieve significant total solute removal while targeting specific ions of interest could have broad implications for water remediation, heavy metal ion removal, high-purity water production and wastewater treatment in the pharmaceutical and other industrial applications.

Furthermore, there is a growing need to develop large-scale separation units that can be operated in continuous electro-swing cycles, similar to those required for temperature, pressure (for gas phases) and pH, polarity or ionic strength (liquid systems) swing adsorption operations. The design and scale-up of such units will be guided and facilitated by the quantitative description of the competitive electrosorption process, which includes the development of the concentration fronts, break-through times, current distributions, etc., to enable process optimization and integration.

Here, we develop a physics-based theoretical framework to study the mass transport of different ionic species in a mixture under continuous flow conditions in an electrochemical cell with Faradaically modulated redox active electrodes for selective adsorption applications. The model accounts for thermodynamic driving forces for solute adsorption on the electrodes, and kinetics of the competitive electrosorption reactions, coupled with mass transport phenomena (diffusion, convection and electromigration) within the electrosorption cell. We present the theoretical model with appropriate boundary and initial conditions, and define the dimensionless variables and groups that control the various operation regimes to understand their impact on the competitive electrosorption processes; a smaller set of the dimensionless groups is derived based on the underlying physics and on a time scale analysis of the processes. The range of values over which these groups must lie to ensure operation in the desired processing regime to meet the process targets are delineated. In the results section, simulation results for three case studies are presented, which cover a wide range of inlet steam conditions that are of practical interest.

**Theory**

The general concepts of the Faradaically modulated selective sorption process are illustrated schematically in Figure 1. The electrosorption cell adapts the common flow-between architecture, in which the ionic mixture in aqueous solution enters the cell from the left, passes over the surfaces of the electrode pair and exits to the right, with an electric field applied across the cell. The electrodes are modified with redox active materials and the active area for ion adsorption is therefore the electrode surface. To achieve high adsorption capacity, the surface-to-volume ratio of an electrosorption cell is usually very large, so the thickness of the channel is on the order of tens to hundreds of microns, which falls into the category of microfluidic devices [48–50]. Since bulk flow direction is perpendicular to the primary direction of ion electromigration and diffusion, at least a two-dimensional model is necessary to capture the coupling of advection and the orthogonal diffusive transport and electromigration [51,52]. We assume that the width of the cell (into the page) is much larger than the height of the channel, as is typical in an ion adsorption cell, and thus that the flow is essentially uniform across the width of the electrode. This allows us to avoid the complexity of a three dimensional velocity field, and to use a width-averaged velocity with a fully developed parabolic profile (Poiseuille flow) between the electrodes [52].

The cathode (at the top) is grounded such that its electrostatic potential at the current collector is always zero, and the electrostatic potential at the anode (at the bottom) equals the applied voltage across the cell. The system of particular interest contains a ternary mixture, where the target anions $A^-$ and supporting anions $X^-$ compete for surface binding sites on the anode and the counter cations $Y^+$ are adsorbed onto the cathode to maintain electrolyte electroneutrality. During the adsorption step, a positive voltage (relative to the equilibrium voltage, as defined later) is applied across the cell to activate the immobilized redox active species on the electrodes as the ionic mixture fills the channel. Due to the different binding affinities of the target anions and the supporting anions with the redox active moieties on the anode, the concentration wave fronts for the two components will move at different velocities, indicated by the two dashed lines. Under typical conditions in which the supporting anions move faster than the target anions, the electrode surface will first be activated and stabilized electrostatically by the supporting anions. When the target anions propagate along the bed length, they displace the supporting anions and form more stable surface complexes with the redox active moieties on the electrode surface. During the desorption step, the applied voltage is decreased such that the surface redox moieties are restored to their original uncharged states, and the adsorbed ions are released to the stripping stream. Note that the theoretical framework developed here is sufficiently general to be easily extended to systems with more than three ionic species.

**Model formulation**

**Governing equations**

The general conservation equations of ion concentrations take the following form, with no homogenous reaction in the bulk:

$$\frac{\partial c_i}{\partial t} = -\vec{\nabla} \cdot \vec{N}_i \tag{1}$$

where $c_i$ and $\vec{N}_i$ are molar concentration and total flux of ion $i$ ($i=A,X,Y$), respectively, and $t$ is time. The flux for each ionic species follows the Nernst Planck equation [53] with contributions from diffusion, convection and electromigration under dilute electrolyte assumptions and unit activity coefficients [54]:

$$\vec{N}_i = -D_i \vec{\nabla} c_i - D_i \frac{z_i F}{RT} c_i \vec{\nabla} \phi + \vec{u} c_i \tag{2}$$

where $D_i$ is the diffusion coefficient for species $i$, $\phi$ is the electrostatic potential, $F$ is the Faraday constant, $R$ is the ideal gas constant, and $T$ is the temperature of the system, which is set to be room temperature.

Fully developed laminar flow is assumed within the channel, i.e. $u(y) = \frac{6U}{H^2}y(H-y)$, where $H$ is the height of the channel and $U$ is the height-averaged velocity in the axial direction. The governing equations for mass transport of each ionic species $i$ ($i=A,X,Y$), are then given by:

$$\frac{\partial c_i}{\partial t} = \frac{\partial}{\partial x}\left(D_i \frac{\partial c_i}{\partial x}\right) + \frac{\partial}{\partial y}\left(D_i \frac{\partial c_i}{\partial y}\right) - u(y)\frac{\partial c_i}{\partial x} + \frac{\partial}{\partial x}\left(D_i \frac{z_i F}{RT} c_i \frac{\partial \phi}{\partial x}\right) + \frac{\partial}{\partial y}\left(D_i \frac{z_i F}{RT} c_i \frac{\partial \phi}{\partial y}\right) \quad (3)$$

The electrostatic potential is defined implicitly by the coupled current conservation law:

$$\frac{\partial \rho_e}{\partial t} = -\vec{\nabla}\cdot\vec{i} \quad (4)$$

Electroneutrality is assumed throughout the bulk electrolyte, usually a good assumption beyond a Debye length from the electrode (~5 nm for a 10 mM binary monovalent electrolyte), which is significantly smaller than the height of the microfluidic channel (on the order of 100 um). Therefore, the total charge density $\rho_e = \sum_i z_i c_i = 0$, and Equation (4) is simplified to $\vec{\nabla}\cdot\vec{i} = 0$, where the ionic current has contributions from all ionic species in the cell,

$$\vec{i} = \sum_i z_i F \vec{N}_i = \sum_i z_i F\left(-D_i \vec{\nabla} c_i - D_i \frac{z_i F}{RT} c_i \vec{\nabla}\phi + \vec{u}c_i\right) \quad (5)$$

**Reaction mechanism**

The competitive electrosorption at the anode between the target anions $A^-$ and the supporting anions $X^-$ is described by the reaction network given in Figure 2. Based on earlier studies on electrochemically mediated adsorption [14,27,55], the current contribution from capacitive charging can be assumed to be negligible compared with that associated with the Faradaic process, and thus the ionic fluxes from capacitive charging are neglected in this study.

The reaction network contains three reactions, explicitly stated in Equation (6), two of which are Faradaic processes and the third a non-Faradaic ion-exchange step. The reaction rates of the three reactions are denoted by $R_{F,X}$, $R_{F,A}$ and $R_{ad}$ respectively. The two Faradaic reactions at the anode are further classified as electrosorption reactions [22], where the immobilized redox active center on the electrode surface undergoes charge transfer and loses an electron, and an anion is adsorbed simultaneously to maintain electroneutrality on the surface.

$$\begin{aligned} R^* + X^- &\rightleftharpoons O^+ X^- + e^- \\ R^* + A^- &\rightleftharpoons [O^+ A^-] + e^- \\ O^+ X^- + A^- &\rightleftharpoons [O^+ A^-] + X^- \end{aligned} \quad (6)$$

The surface coverage of different (oxidation) states is described by a lattice (ideal gas mixture) model, $\theta_i \equiv c_{s,i}/c_{s,0}$, $i=R,A,X$, in which the asterisk on $R$ indicates the unbounded surface sites under the reduced state and it is assumed that there are no interactions between the surface sites under all conditions. $c_{s,0}$ is the total surface site density and the surface coverages sum to unity. Initially, most of the surface species at the anode are in the uncharged, reduced form $R^*$. After a positive polarization is applied, the surface species are oxidized and pair with the supporting or target anions ($X^-$ or $A^-$), to form $O^+X^-$ or $[O^+A^-]$, respectively, due to electrostatic attraction. Square brackets are used to indicate that the formation of $[O^+A^-]$ is more favorable than the formation of $O^+X^-$, as the former complex is thermodynamically more stable than $O^+X^-$ due to molecularly specific interactions [14]. In addition to the electrosorption reaction, $[O^+A^-]$ can also be formed through ion-exchange of $A^-$ with $X^-$ on the complex $O^+X^-$. The concentrations of surface species are denoted by $c_{s,R}$, $c_{s,X}$ and $c_{s,A}$. Since the surface species are immobilized at the electrode surface, there is no flux contribution from mass transport within and outside of the electrode, and the reaction kinetics are simply described by:

$$\frac{\partial c_{s,R}}{\partial t} = -\left(R_{F,X} + R_{F,A}\right)$$
$$\frac{\partial c_{s,X}}{\partial t} = R_{F,X} - R_{ad} \quad (7)$$
$$\frac{\partial c_{s,A}}{\partial t} = R_{F,A} + R_{ad}$$

The electrosorption reaction rates are derived from generalized Butler-Volmer kinetics, with the current density given by $i_{F,i} = nFR_{F,i}, i = A, X$, which is positive for oxidation, or anodic current; $n=1$ for the single electron transfer reaction and $R_{F,i}$ indicates the reaction rate. The detailed derivation of $R_{F,X}$ from a non-equilibrium thermodynamics perspective [54] is included in Supporting Information S2.1. At equilibrium, the current density is zero, and thus, if we assume the activity of electrons $a_e = 1$, i.e., $a_{s,X} \equiv \frac{\theta_X}{\theta_R^*}$, the equilibrium voltage is:

$$\Delta\phi_{eq,X} = \Delta\phi_{eq,X}^\Theta + \frac{RT}{F}\ln\left(\frac{a_{s,X}a_e}{a_X}\right) = \Delta\phi_{eq,X}^\Theta + \frac{RT}{F}\ln\left(\frac{\theta_X}{\theta_R^* a_X}\right) \quad (8)$$

which recovers the Nernst equation. $a_X$ and $a_{s,X}$ are the activities of the mobile anion and the immobilized surface complex with the oxidized state, respectively. The current density can be further written in terms of overpotential $\eta_X$ as

$$i_{F,X} = i_{0,X}\left(e^{\frac{(1-\alpha)F\eta_X}{RT}} - e^{-\frac{\alpha F\eta_X}{RT}}\right) \quad (9)$$

$$\eta_X = \Delta\phi - \Delta\phi_{eq,X} \quad (10)$$

$$i_{0,X} = Fc_{s,0}k_X^0 \frac{1}{\gamma_\ddagger} a_X^\alpha \left(a_{s,X}a_e\right)^{1-\alpha} \quad (11)$$

The activity coefficient of the transition state is defined as $\gamma_\ddagger \equiv \frac{1}{\theta_R^*}$, which is assumed to exclude a single reduced (unbound) state, and indicates the increase in the activation energy with fewer unbound sites [56]. $k_X^0$ denotes the rate constant and $\alpha$ the charge transfer coefficient for the Faradaic reaction. With the expressions for $\gamma_\ddagger$ and $a_{s,X}$, the thermodynamically consistent exchange current density for generalized Butler-Volmer kinetics is given by $i_{0,X} = Fk_{0,X}(\theta_R^* a_X)^\alpha \theta_X^{1-\alpha}$. The prefactor $k_{0,X} = k_X^0 c_{s,0}$, and can usually be extracted experimentally from rotating disk electrode measurements or linear sweep voltammetry, i.e., $k_{0,X} = \frac{k_{0,X}^{\text{exp}}}{c_{s,0}^{\text{exp}}} c_{s,0}$ [57,58], where $k_{0,X}^{\text{exp}}$ and $c_{s,0}^{\text{exp}}$ denote the experimentally measured reaction rate and total surface site density, respectively.

The kinetics of the second electrosorption reaction involving the target anions are derived similarly, and results are represented by simply replacing the subscript $X$ with $A$ throughout equation (8) to (11), with details given in SI S2.2.

The surface ion-exchange rate is described by second order reaction kinetics, which also leads to the following thermodynamically consistent formulation [54] (see SI S2.3):

$$R_{\text{ad}} = k_f c_A c_{s,X} - k_b c_{s,A} c_X = k_f c_{A,0} c_{s,0} \left( \tilde{c}_A \theta_X - K_d \theta_A \tilde{c}_X \right) \tag{12}$$

where $K_{\text{ad}}(K_d)$ are the adsorption (desorption) equilibrium constants of the ion exchange reaction, $K_d = \frac{k_b}{k_f} = \frac{1}{K_{\text{ad}}}$.

The driving force for the ion-exchange step is the difference in surface adsorption energy between the target anions $\Delta G_A^\Theta$ and the supporting anions $\Delta G_X^\Theta$ under standard state, which is denoted as $\Delta E_{\text{ad}} = \Delta G_A^\Theta - \Delta G_X^\Theta$. The equilibrium constant for this reaction is then $K_{\text{ad}} = e^{-\Delta \tilde{E}_{\text{ad}}}$, where $\Delta \tilde{E}_{\text{ad}} = \frac{\Delta E_{\text{ad}}}{RT}$. If we define dimensionless potential as $\tilde{\phi} = \frac{F}{RT}\phi$, and note that $\Delta \phi_{\text{eq}}^\Theta = \frac{\Delta G^\Theta}{F}$, then,

$$\Delta \tilde{\phi}_{\text{eq},A}^\Theta - \Delta \tilde{\phi}_{\text{eq},X}^\Theta = \Delta \tilde{E}_{\text{ad}} = -\ln(K_{\text{ad}}) \tag{13}$$

and thus, the surface adsorption energy $\Delta E_{\text{ad}}$ or equivalently the equilibrium constant of the ion exchange reaction $K_{\text{ad}}$, links the two Faradaic reactions.

At equilibrium, $R_{\text{ad}}$ equals zero, and the surface coverage ratio for the two competing anions is,

$$\frac{\theta_{A,\text{eq}}}{\theta_{X,\text{eq}}} = \frac{1}{K_d} \frac{\tilde{c}_{A,\text{eq}}}{\tilde{c}_{X,\text{eq}}} \simeq \frac{1}{K_d \gamma} = \frac{K_{\text{ad}}}{\gamma} \tag{14}$$

which depends on the final concentration ratio of the supporting ions versus the target ions, roughly equal to the inlet concentration ratio $\gamma = c_{X,\text{in}}/c_{A,\text{in}}$. Under general inlet conditions, the selectivity of the process can be defined as,

$$S_{A/X} = \frac{c_{X,\text{eq}}}{c_{A,\text{eq}}} \frac{\theta_{A,\text{eq}}}{\theta_{X,\text{eq}}} \simeq K_{\text{ad}} \tag{15}$$

which only depends on the thermodynamics metrics, i.e., the adsorption energy ($\Delta \tilde{E}_{\text{ad}}$) of the target anions over the supporting anions on the oxidized electrode surface, under the ideal lattice gas model (Langmuir isotherm).

**Boundary conditions**

The boundary conditions of the system are shown schematically in Figure 3. Specifically, for $x = 0$, the inlet concentrations are specified using the Dirichlet boundary condition, $c_i = c_{i,0}$, $i = A, X, Y$. For $x = L$, it is assumed that the ionic flux is purely due to the convection of the chemical species along the stream, and that axial diffusion is negligible, $\frac{\partial c_i}{\partial x} = 0$.

For $y = 0$, the flux of each chemical species is related to the rate of adsorption, and with $\vec{n}$ as the unit normal vector pointing outward, we have

$$\vec{n} \cdot \vec{N}_A = R_{F,A} + R_{ad}$$
$$\vec{n} \cdot \vec{N}_X = R_{F,X} - R_{ad} \quad (16)$$
$$\vec{n} \cdot \vec{N}_Y = 0$$

For $y = H$, the fluxes of both anions are zero, and the flux of cations carries the current to the cathode to close the circuit of the electrosorption cell. The cathode is grounded such that $\phi_e = 0$.

$$\vec{n} \cdot \vec{N}_A = 0$$
$$\vec{n} \cdot \vec{N}_X = 0 \quad (17)$$
$$\vec{n} \cdot \vec{N}_Y = -R_{F,Y}$$

The Faradaic reaction rate is described by Butler-Volmer kinetics as well, where

$$R_{F,Y} = \frac{i_{0,Y}}{F}\left(e^{\frac{(1-\alpha)F\eta_Y}{RT}} - e^{-\frac{\alpha F\eta_Y}{RT}}\right)$$
$$\eta_Y = \Delta\phi - \Delta\phi_{eq} = \phi_e - \phi - \Delta\phi_{eq,Y} \quad (18)$$
$$i_{0,Y} = Fk_{0,Y}a_Y^{1-\alpha}$$

For the cathode reaction, we have made two further assumptions. Firstly, the capacity of the cathode is relatively large, and therefore the charging and discharging processes do not alter the voltage from its equilibrium value significantly, i.e., $\Delta\phi_{eq,Y} \approx \Delta\phi_{eq,Y}^{ref}$. Furthermore, fast kinetics at the cathode are assumed in comparison with the kinetics at the anode, i.e., $k_{0,Y} \gg k_{0,X}, k_{0,A}$. Under these two assumptions, which are close to setting $\phi(y = H) = 0$, it is ensured that the cathode reaction is not the limiting factor in the process, and allows us to focus on the effect of the competitive electrosorption kinetics at the anode. These restrictions can be relaxed by including more realistic kinetics of the cathode material in any future study.

The boundary conditions for the electrostatic potential are specified as follows: at $x = 0, x = L$, there is no diffusive flux or potential gradient, the convective flux is also zero due to bulk electroneutrality, and therefore no current flows in the horizontal direction at the inlet and outlet, i.e., $\vec{n} \cdot \vec{i} = 0$; at $y = 0$, the current density in the vertical directions equals the sum of ionic fluxes, $\vec{n} \cdot \vec{i} = -F(R_{F,A} + R_{F,X})$; and at $y = H$, the cathodic Faradaic reaction rate balances the anodic reaction rate, $\vec{n} \cdot \vec{i} = -FR_{F,Y}$.

**Initial conditions and consistent initialization**

It is assumed that initially there is dilute amount of supporting electrolyte within the channel,

$$c_A(x,y,t=0)=0$$
$$c_X(x,y,t=0)=c_{X,\text{init}} \quad (19)$$
$$c_Y(x,y,t=0)=c_{X,\text{init}}$$

with a concentration that is of the same magnitude as that of the target anions in the inlet mixture stream, i.e., $c_{X,\text{init}}=c_{A,0}$, in the micromolar range. This avoids dealing with ill-posed equations under zero concentrations and allows faster numerical convergence.

The initial surface coverages of different oxidation states are specified to be

$$\theta_R^*(t=0)=\theta_{R,\text{init}}^*$$
$$\theta_X(t=0)=\theta_{X,\text{init}} \quad (20)$$
$$\theta_A(t=0)=0$$

where $\theta_{R,\text{init}}^*$ and $\theta_{X,\text{init}}$ are constants. In the simulation, $\theta_{X,\text{init}}=0.0099$.

Since the electrostatic potential is solved implicitly by the current conservation law, the differential-algebraic PDE system requires consistent initialization to ensure the system of equations is well-posed. The initial cell voltage is set to be the same as the equilibrium voltage of the adsorption step, $V_{\text{cell,init}}=V_{\text{cell,ad}}=\Delta\phi_{\text{eq},X}^{\text{ref}}-V_T\ln(\gamma)$, and the other set of equations to be satisfied under the initial conditions are summarized in the Supporting Information S6.1; the dimensionless form is presented in SI S1.2 equation (18).

This set of equations can be solved to get consistent initialization for all the implicit variables, including the electrostatic potential at the anode (denoted by $\phi_{a,\text{init}}$), the non-zero initial current and overpotential, etc. The initial condition for the electrostatic potential is specified to be

$$\phi(t=0)=(1-y/H)\phi_{a,\text{init}} \quad (21)$$

which has a linear profile in the vertical direction and is uniform along the axial direction. Another option for initialization of the ion adsorption cell is to assume that the cell starts at rest, with an initial current of zero, and the initial cell voltage is solved for consistently from the same set of equations as in S1.2 equation (18). Since the two initialization methods arrive the same steady state results, the first approach is selected and reported in this work for simplicity, and we suggest readers refer to SI S6.2 for detailed discussion for the second approach of consistent initialization if interested.

**Dimensionless variables and groups**

We nondimensionalize the equations to reduce the number of free parameters and show the important dimensionless groups that govern the complex competitive adsorption behavior of ionic species under the electric field. The dimensionless variables are defined as,

$$\tilde{c}_A = \frac{c_A}{c_{i,\text{ref}}}, \ \tilde{c}_X = \frac{c_X}{c_{i,\text{ref}}}$$

$$\theta_R^* = \frac{c_{s,R}}{c_{s,0}}, \ \theta_X = \frac{c_{s,X}}{c_{s,0}}, \ \theta_A = \frac{c_{s,A}}{c_{s,0}}$$

$$\tilde{y} = \frac{y}{H}, \ \tilde{x} = \frac{x}{H}, \ \hat{x} = \frac{x}{PeH}$$

$$\tilde{t} = \frac{t}{t_d}, \ \hat{t} = \frac{t}{t_c} = \frac{\tilde{t}}{Gz} \quad (22)$$

$$\tilde{\phi} = \frac{F}{RT}\phi, \ \tilde{\eta} = \frac{F}{RT}\eta$$

$$\tilde{N}_i = \frac{N_i}{\frac{D_{\text{ref}}c_{i,\text{ref}}}{H}}, \ \tilde{R}_i = \frac{R}{\frac{D_{\text{ref}}c_{i,\text{ref}}}{H}}, \ \tilde{i} = \frac{i}{\frac{FD_{\text{ref}}c_{i,\text{ref}}}{H}}$$

where the reference concentration used to normalize the bulk concentration is the inlet concentration of the target anions, i.e., $c_{i,\text{ref}} = c_{A,0}, i = A, X, Y$, and the reference diffusion coefficient is that of the target anions $D_{\text{ref}} = D_A$. The surface coverage percentages of the different states are nondimensionalized by the total surface site density, voltages are normalized by the thermal voltage $RT/F$, and fluxes and current densities are normalized by the diffusive flux of the target anions under the reference conditions.

Since the aspect ratio of the ion adsorption channel, i.e. $L/H$ is usually very large, the length of the channel is further normalized by the (vertical) Peclet number, which is defined below in equation (23). Thus the rescaled length of the bed is $\hat{L} = \frac{L/H}{Pe} = \frac{t_c}{t_d} = Gz$. The Graetz number ($Gz$) indicates the ratio of time scales between convection in the axial direction versus diffusion across the height of the cell. For easier comparison of simulation results irrespective of the flow rate ($Pe$), an additional scaling factor $a$ is used to renormalize the length of the channel to be 10, i.e, $a = \frac{10}{\hat{L}} = \frac{10}{Gz}$, such that the geometry of the simulation domain has a fixed aspect ratio of 10:1. Inevitability, $a$ will appear in the governing equations and boundary conditions whenever the axial distance is involved. For simplicity and for clarity of the equations, $a$ is not shown in the derivations.

The important dimensionless groups that appear naturally when the equations are cast in the non-dimensional form, are:

$$Pe = \frac{UH}{D_{\text{ref}}}, \quad \beta = \frac{L}{H}, \quad Gz = \frac{\beta}{Pe}, \quad \gamma = \frac{c_{X,0}}{c_{A,0}}$$

$$K_{\text{ad}} = \frac{1}{K_d} = \frac{k_f}{k_b}, \quad \varepsilon = \frac{c_{A,0} H}{c_{s,0}}, \quad Da = \frac{k_f c_{s,0} H}{D_{\text{ref}}}$$

$$\nu_i = \frac{k_{0,i}^{\text{ref}}}{\frac{D_{\text{ref}} c_{i,\text{ref}}}{H}} = \frac{k_{0,i}}{\frac{D_{\text{ref}} c_{i,\text{ref}}}{H}} \left(\frac{c_{i,\text{ref}}}{c^{\Theta}}\right)^{\alpha}, \quad i = A, X$$

$$\tilde{U}_{\text{des}} = \frac{U_{\text{des}}}{U}, \quad \gamma_{i,\text{wash}} = \frac{c_{i,\text{wash}}}{c_{i,\text{ref}}}, \quad \gamma_{i,\text{des}} = \frac{c_{i,\text{des}}}{c_{i,\text{ref}}}$$

$$\Gamma_{\text{ad},i} = \frac{\theta_{i,\text{eq,ad}}}{\theta_{R,\text{eq,ad}}^{*}}, \quad \Gamma_{\text{des},i} = \frac{\theta_{i,\text{eq,des}}}{\theta_{R,\text{eq,des}}^{*}}$$

$$\Delta \tilde{\phi}_{\text{eq},i}^{\text{ref}} = \Delta \tilde{\phi}_{\text{eq},i}^{\Theta} + \ln\left(\frac{c^{\Theta}}{c_{i,\text{ref}}}\right), \quad \Delta \tilde{\phi}_{\text{eq},A}^{\text{ref}} = \Delta \tilde{\phi}_{\text{eq},X}^{\text{ref}} - \ln(K_{\text{ad}})$$

$$\tilde{V}_{\text{cell,ad}} = \Delta \tilde{\phi}_{\text{eq},i}^{\text{ref}} + \ln\left(\frac{\Gamma_{\text{ad},i}}{\tilde{c}_{i,\text{eq,ad}}}\right), \quad \tilde{V}_{\text{cell,des}} = \Delta \tilde{\phi}_{\text{eq},i}^{\text{ref}} + \ln\left(\frac{\Gamma_{\text{des},i}}{\tilde{c}_{i,\text{eq,des}}}\right) \quad (23)$$

$$\Delta \tilde{\phi}_{\text{ad}} = \tilde{V}_{\text{cell,ad}} - \Delta \tilde{\phi}_{\text{eq},A}^{\text{ref}}, \quad \Delta \tilde{\phi}_{\text{des}} = \tilde{V}_{\text{cell,des}} - \Delta \tilde{\phi}_{\text{eq},A}^{\text{ref}}$$

In addition to the Peclet number ($Pe$), aspect ratio ($\beta$), Graetz number ($Gz$), inlet concentration ratio ($\gamma$) and equilibrium constant for the adsorption reaction ($K_{\text{ad}}$) introduced above, $\varepsilon$ denotes the surface adsorption capacity relative to the bulk concentration, $\nu_i$ and $Da$ denote the dimensionless reaction rate of the Faradaic electrosorption reactions and the ion exchange reaction, respectively, $\tilde{U}_{\text{des}}$ is the ratio of flow rate during the desorption step to that in the adsorption step, and $\gamma_{i,\text{wash}}$ and $\gamma_{i,\text{des}}$ denote the inlet concentration of anions during the wash and desorption steps relative to the reference concentrations. $\Gamma_{\text{ad},i}$ and $\Gamma_{\text{des},i}$ are the equilibirum surface coverage ratios of anion $i$ versus those in the unbound reduced state. $\tilde{V}_{\text{cell,ad}}$ and $\tilde{V}_{\text{cell,des}}$ are the applied cell voltage during adsorption and desorption steps, respectively; $\Delta \tilde{\phi}_{\text{ad}}$ and $\Delta \tilde{\phi}_{\text{des}}$ are the applied voltage in reference to the equilibrium voltage of the electrosorption reaction of the target anions under the reference concentration, i.e, $\Delta \tilde{\phi}_{\text{eq},A}^{\text{ref}}$, which is typically a material specific property. The dimensionless governing equations, boundary conditions and consistent set of initial conditions can be found in the supporting information S1.

**Boundary condition for the three-step stop-flow operation**

For the adsorption and desorption steps, the electrosorption system follows the same governing equations, and only certain operational variables are changed as reflected in the boundary conditions. For the desorption step, the applied voltage is reduced to allow the release of captured anions to the stripping stream. At the same time, the flow rate is reduced such that a higher concentration of the target anions results in the regeneration stream, which is beneficial for downstream processing.

Besides the common two-step adsorption/desorption operation, we also add an intermediate wash step during which the applied voltage is held and the ion adsorption cell is washed with a stripping stream. Two possible stripping strategies are considered for the wash and desorption steps:
(1) wash the cell with dilute supporting electrolyte ($\gamma_{X,\text{wash}} = 1, \gamma_{A,\text{wash}} = 0$), and desorb the captured target anions to the same stripping solution ($\gamma_{X,\text{des}} = 1, \gamma_{A,\text{des}} = 0$);
(2) wash the cell with a high concentration solution of pure target anions ($\gamma_{X,\text{wash}} = 0, \gamma_{A,\text{wash}} = 2$), and desorb the captured target anions into the same stripping solution ($\gamma_{X,\text{des}} = 0, \gamma_{A,\text{des}} = 2$).

In both approaches, the concentrated supporting ions in the adsorption cell at the end of the adsorption step are washed away, and the surface coverage by the supporting anions drops to nearly zero after the wash step; both effects synergistically enable the release of captured anions to the stripping stream with a higher purity. The conditions for average flow velocity, inlet concentrations and cell voltages are summarized in Table 1, and a schematic of the three-step stop-flow operation with the second wash-desorption strategy can be found in SI (Figure S2).

The numerical simulation was conducted in COMSOL Multiphysics 5.3a, a commercial finite element solver. Details about the numerical implementation and meshing of the simulation domain can be found in SI S1.3 Figure S1. The simulation time for the adsorption step was set to be slightly longer than the propagation time scale $\tilde{t}_p$ to obtain a full break-through of the target anions. The simulation time for the wash step was 2 bed volumes, and the simulation time for the desorption step was set to 1 or 2 bed volume(s) depending on whether the stop-flow step was implemented. Parameter values used in the simulation were selected over the proper range of interest and are summarized in Table 2; the derived values of the dimensionless groups are also shown in Table 3.

**Discussion of design space and operation regimes**

**Choice of dimensionless groups**

The transport of ionic species and the competitive surface electrosorption processes within the ion adsorption cell are controlled by the dimensionless groups defined in Table 3. Determination of the minimal set of independent dimensionless groups and an understanding of their impact on the process is crucial to achieve the desired separation performance. Some of the dimensionless groups are set by process specification, such as the inlet concentration ratio of the two competing anions $\gamma$, which depends on the contaminated source water and cannot be changed easily; some are process variables, which can be tuned by changing the process set points, such as flow rate, applied cell voltage, etc; while some depend on the material properties, such as the adsorption equilibrium constant, reaction rate of the electrosorption reaction, etc, which are independent of process variations. Here, $Gz$, $\varepsilon$, $\Gamma_{\text{ad},A}$, $\Gamma_{\text{ad},X}$ and $\Gamma_{\text{des},A}$ are chosen as the independent dimensionless process specifications that constrain the process inputs within a certain range; $K_{\text{ad}}$, $Da$ and $\nu_i$ are chosen as the independent material properties. The physical meanings of these groups can be understood as follows.

The Graetz number $Gz$ controls the time required for ionic species to move from the inlet to the outlet relative to the characteristic diffusion time in the vertical direction. When $Gz \ll 1$, due to a fast flow rate or a short bed length, convection dominates over diffusion and a mass transfer boundary layer will be formed near the surface of the electrodes, so that the device is operated under "entrance region" conditions. This regime is typically chosen in studies of the surface reaction kinetics in biochemical assays by surface plasma reflection techniques [59–61] or in flow batteries to achieve a high current density [52,62]. However, operating in the entrance region leads to a low capture ratio for the target ions, which is unfavorable when the samples are small in volume or very expensive, such as in single cell analysis [63]. For water treatment applications in particular, where it is sometimes desirable to recover high-valued ions from a mixture stream [64,65] or to remove certain toxic ions from the source water [14,17] for safe disposal, a high capture fraction of the target ions is an important metric for the water treatment process. In this case, an ion adsorption cell with a long capture bed is preferred, such that the target ions have significant time to diffuse across the height of the channel before they exit the cell, i.e., operation in the "fully developed region" of forced convection. Therefore, in this study we only focus on the cases where $Gz \gg 1$.

The bulk-to-surface capacity ratio $\varepsilon$ determines the degree of depletion of the target ions in the bulk by specifying the ratio of the bulk concentration of target anions with respect to the total surface site density. When $\varepsilon \ll 1$, the amount of target anions in the ion adsorption cell is much smaller than the surface capacity of the electrodes. Therefore, the target ions will be depleted initially, and break-through of the ions will not occur until the surface is fully saturated with the bound ions, in behavior reminiscent of that of a packed bed adsorption column. On the contrary, when $\varepsilon \gg 1$, i.e. the concentration of target anions is much larger than the available binding sites on the electrode surface, the target ions will quickly saturate the surface of the ion adsorption cell and flow out of the channel with little or no retention, resulting in a low capture ratio of the target ions. Therefore, in the simulation, we are interested in the case where $\varepsilon \ll 1$, such that the target ions will be depleted in the channel initially, and allows a separation of the concentration wave fronts for the supporting and target anions to achieve selective removal of the target ions from the mixture stream.

The other three process specifications are $\Gamma_{\text{ad},A}$, $\Gamma_{\text{ad},X}$ and $\Gamma_{\text{des},A}$ which denote the equilibrium surface coverage ratios between the target or supporting anions and the unbound reduced states at the end of adsorption or desorption steps. In order to achieve higher selectivity towards target anions, it is desired to have $\Gamma_{\text{ad},A} \gg 1$ and $\frac{\Gamma_{\text{ad},A}}{\Gamma_{\text{ad},X}} \gg 1$ after the adsorption step reaches equilibrium; similarly, it will be preferred to have $\Gamma_{\text{des},A} \ll 1, \Gamma_{\text{des},X} \ll 1$ at the end of the desorption step. In this study, $\Gamma_{\text{ad},i}$ and $\Gamma_{\text{des},A}$ are assigned fixed values, i.e, $\Gamma_{\text{ad},A} = e^{\Delta\tilde{\phi}_{\text{ad}}} \tilde{c}_{A,\text{eq,ad}} \gg 1$, $\Gamma_{\text{ad},X} = K_d e^{\Delta\tilde{\phi}_{\text{ad}}} \tilde{c}_{X,\text{eq,ad}} \approx 1$, $\Gamma_{\text{des},A} = e^{\Delta\tilde{\phi}_{\text{des}}} \tilde{c}_{A,\text{eq,des}} \ll 1$ (see Table 3); under these conditions, the surface coverage ratio of target to supporting anions is also maintained at a large value during adsorption, $\frac{\theta_{A,\text{eq,ad}}}{\theta_{X,\text{eq,ad}}} = K_{\text{ad}} \frac{\tilde{c}_{A,\text{eq,ad}}}{\tilde{c}_{X,\text{eq,ad}}} \approx \frac{K_{\text{ad}}}{\gamma} \gg 1$ and the coverage ratio of supporting anions to the unbound state is maintained low, $\Gamma_{\text{des},X} = K_d \Gamma_{\text{des},A} \frac{\tilde{c}_{X,\text{eq,des}}}{\tilde{c}_{A,\text{eq,des}}} \ll 1$. Detailed derivations can be found in SI S2.4.

It is important to study the validity of this technology for a range of inlet concentration ratios ($\gamma$) between the target and supporting anions for robustness and process scale-up consideration. Below in Section 4, we explore the three scenarios of $\gamma \gg 1, \gamma = 1,$ and $\gamma \ll 1$ in more detail with fixed target anion inlet concentration in the micromolar range through three case studies. For the case studies, the process inputs ($Gz$, $\varepsilon$, $\Gamma_{ad,A}$, $\Gamma_{ad,X}$ and $\Gamma_{des,A}$) were chosen to cover the preferred operational ranges and maintained at fixed values throughout the studies.

An important insight from the analysis above is that the material selection requirement of the redox active material can be relaxed accordingly for smaller inlet concentration ratios between the two competing anions as long as $\frac{\theta_{A,eq,ad}}{\theta_{X,eq,ad}} \approx \frac{K_{ad}}{\gamma} \gg 1$; i.e., the surface adsorption energy (i.e. $K_{ad}$) of the electrode materials can be reduced proportionally for smaller $\gamma$ with maintainance of the same equilibrium surface coverage ratios between $\theta_{A,eq,ad}$, $\theta_{X,eq,ad}$ and $\theta^*_{R,eq,ad}$. In order to facilitate comparison between the scenarios with various inlet concentration ratios, the dimensionless kinetic parameters, (i.e. $Da$ and $\nu_i$) for the various redox active materials were assumed to be constant under the three scenarios for all three case studies; furthermore, it was assumed that the forward rate constant $k_f$ of the various redox active material was fixed, and the backward rate constant $k_b$ for the ion-exchange reaction of the electrode materials was adjusted to accommodate various $K_{ad}$ values. Since $\tilde{V}_{cell,ad} = \Delta\tilde{\phi}^{ref}_{eq,X} + \ln(\Gamma_{ad,X}/\tilde{c}_{X,eq,ad}) \approx \Delta\tilde{\phi}^{ref}_{eq,X} + \ln(\Gamma_{ad,X}/\gamma)$ (see SI S2.4), the applied cell voltage should also change with the inlet concentration ratio to keep the equilibrium surface coverage of the anions at the end of the adsorption step similar between the different scenarios. $\tilde{V}_{cell,des} = \tilde{V}_{cell,ad} - \ln\left(\frac{\Gamma_{ad,A}}{\Gamma_{des,A}}\right) - \ln\left(\frac{\tilde{c}_{A,eq,des}}{\tilde{c}_{A,eq,ad}}\right) \approx \tilde{V}_{cell,ad} - 8$ (see SI S5.2). Therefore, under various inlet concentration ratios between the two competing anions, both the electrode material selection criteria (i.e. $K_{ad}$) and the process voltage input (i.e. $\tilde{V}_{cell,ad}$ and $\tilde{V}_{cell,des}$) were adjusted to fulfill the process specifications assigned above; meanwhile $\Delta\tilde{\phi}_{ad}$ and $\Delta\tilde{\phi}_{des}$ was essentially kept constant despite the variation of $\gamma$ (see Table 3).

**Time scale analysis**

Although the dimensionless groups discussed in equation (23) lump the input parameters of the system into a more compact set, the large number of dimensionless groups still renders the problem a high-dimensional design space, which is challenging to explore and optimize. It would be beneficial to look into the key physical steps within the competitive adsorption processes, and derive criteria to meet requirements of the applications, such as high separation efficiency, to further reduce the design space.

For selective removal of the target anions, the time dependent response of the electrosorption system is governed by the following five time scales: (i) diffusion across the height of the channel, (ii) convection from the inlet to the outlet, (iii) reaction time for the surface electrosorption process, (iv) saturation time for the surface (electro)-adsorption reactions, and (v) wavefront propagation time (of target anions) in the electrosorption cell due to stronger interaction with the electrode surface. For competitive adsorption as studied here, the reaction and saturation time scales have contributions from both the Faradaic electrosorption and the ion-exchange reaction, where the detailed derivations can be found in SI S3.1 and S3.2 respectively. The five time scales can be written as

$$\text{Diffusion}: t_d = \frac{H^2}{D}$$

$$\text{Convection}: t_c = \frac{L}{U}$$

$$\text{Reaction}: t_r = \frac{Hc_{A,0}}{k_f c_{s,0} c_{A,0} + k_{0,A}^{\text{ref}} e^{(1-\alpha)\Delta\tilde{\phi}_{\text{ad}}}} = \frac{1}{\frac{1}{t_{r,\text{ad}}} + \frac{1}{t_{r,F,A}}}$$

$$\text{Surface Saturation}: t_{sat} = \frac{1}{\left(k_f c_{A,0} + k_b c_{X,0}\right) + \frac{k_{0,A}^{\text{ref}}}{c_{s,0}}\left(e^{(1-\alpha)\Delta\tilde{\phi}_{\text{ad}}} + e^{-\alpha\Delta\tilde{\phi}_{\text{ad}}}\right)} = \frac{1}{\frac{1}{t_{sat,\text{ad}}} + \frac{1}{t_{sat,F,A}}}$$

$$\text{Adsorption Front Propagation}: t_p = \frac{L}{U_{\text{eff}}}$$

(24)

The effective wave propagation speed, as derived in SI S4, is

$$U_{\text{eff}} = \frac{U}{1 + \frac{1}{\varepsilon\left(e^{-\Delta\tilde{\phi}_{\text{ad}}} \frac{1}{\tilde{c}_{A,\text{eq}}} + K_d \frac{\tilde{c}_{X,\text{eq}}}{\tilde{c}_{A,\text{eq}}} + 1\right)}}$$ (25)

It can be seen that the concentration wave front velocity of the target anions during the adsorption step decreases (increases) monotonically with the adsorption (desorption) equilibrium constant $K_{\text{ad}}$ ($K_d$): the stronger the binding affinity between the target anions and the electrode surface, the slower will the target anion concentration front propagate through the electrosorption cell.

In order to achieve effective separation of the target anions and the supporting anions, the propagation time scale of the target anion concentration front should be the slowest. Upon nondimensionalization of these time scales with respect to the diffusion time scale, combinations of the following important dimensionless groups appear naturally (SI S3.3). By setting $\tilde{t}_r, \tilde{t}_{sat}, \tilde{t}_d, \tilde{t}_c \ll \tilde{t}_p$, and from the specifications discussed in the earlier section, $Gz \gg 1$, $\varepsilon \ll 1$ and $\Gamma_{\text{ad},A} \approx e^{\Delta\tilde{\phi}_{\text{ad}}} \approx \frac{K_{\text{ad}}}{\gamma} \gg 1$, the convection limited wave propagation regime appears whenever

$$Gz\left(Da + \nu_A e^{(1-\alpha)\Delta\tilde{\phi}_{\text{ad}}}\right) \gg 1$$ (26)

is satisfied. From the previous section, we already know that operation under $Gz \gg 1$ is desired. Therefore, the kinetics of either the electrosorption reaction or the ion exchange step should not be particularly sluggish to ensure that the criterion in equation (26) is satisfied. For a pure surface binding process without Faradaically enhanced adsorption, this expression reduces to $Gz\,Da \gg 1$, which agrees well with the result in Gervais et al [49].

## Results and Discussion

We consider the three different cases in which the supporting electrolyte is in significant excess over the target anions ($\gamma = 100$, environmental contaminants), is equal in concentration to the target species ($\gamma = 1$, chemical separations), and is very dilute relative to the target ($\gamma = 0.01$, polishing and up-concentration). The latter case is rarely studied because it is difficult to examine experimentally and suffers from high energy consumption due to the low overall ionic conductivity, but it is an important regime for deionization and water remediation. In summary, the three cases are designed to cover a wide range of practical interest in water treatment applications.

In the three case studies, all process specifications in Equation (23) were kept constant to allow a fair comparison across the different scenarios; the only exceptions were the thermodynamic material property $K_{ad}$ and the applied cell voltage $\tilde{V}_{cell,ad}$ and $\tilde{V}_{cell,des}$: the equilibrium electrode material adsorption constant was set to scale with the initial concentration ratio of the two competing anions, i.e., $K_{ad} = \frac{\Gamma_{ad,A}}{\Gamma_{ad,X}} \frac{\tilde{c}_{X,eq,ad}}{\tilde{c}_{A,eq,ad}} \approx \frac{\Gamma_{ad,A}}{\Gamma_{ad,X}} \gamma$, and the applied cell voltage during the adsorption/desorption step was adjusted accordingly, i.e., $\tilde{V}_{cell,ad} = \Delta\tilde{\phi}_{eq,X}^{ref} + \ln(\Gamma_{ad,X}/\tilde{c}_{X,eq}) \approx \Delta\tilde{\phi}_{eq,X}^{ref} + \ln(\Gamma_{ad,X}/\gamma)$ and $\tilde{V}_{cell,des} = \tilde{V}_{cell,ad} - 8$ such that the same net voltage difference was maintained under various $\gamma$.

## Concentration and Potential Distribution in Flow Channel

The two-dimensional contour plots of the target and supporting anion concentrations and the electrostatic potential at $\hat{t} = 0.5$ are given in Figure 4; animations of the development of these concentration profiles during the adsorption step are available in the supplement materials SII. When the supporting electrolyte is in significant excess over the target anions, ($\gamma = 100$), its concentration front has moved almost to the half-way point of the channel at $\hat{t} = 0.5$, with slight retardation due to the adsorption of a small fraction of the anions introduced to the system. The target anions, on another hand, have not penetrated far from the entrance of the channel as they are removed from the solution by strong interactions with the Faradaically activated anode binding sites; the velocity of the concentration wave front of the target anions, $\tilde{U}_{eff} = \frac{1}{1+\frac{1}{\varepsilon}\theta_A} \sim \varepsilon \ll 1$, is much smaller than the nominal fluid velocity. A non-uniform electric field is established locally at the anode near the concentration front of the supporting anions due to the dynamic stabilization of oxidized sites by the supporting anions via electrostatic interaction as the anion concentration increases in this region with the approach of the adsorption front. The electric field is close to zero both downstream and upstream of the adsorption front of the supporting anions since in these regions the concentrations are unchanging, and the Faradaically activated moieties on the electrode surface are in equilibrium with the solution in which they are in contact, such that no further oxidation occurs, i.e., there is no current flow.

If the supporting anions are present at a much lower concentration than the target anions, i.e., $\gamma = 0.01$, so that the solution is "unsupported", their concentration front moves slowly as the ions are depleted from the solution by binding with the activated redox centers on the electrode surface, and the depletion is significant relative to the amount introduced to the channel. As derived in S4, $\widetilde{U}_{\text{eff},X} = \frac{1}{1+\frac{\theta_X}{\varepsilon\gamma}}$, so that a decrease in supporting anion concentration will reduce the propagation speed of supporting anions in the cell significantly. The magnitude of the electrostatic potential is higher than in the fully supported case, due to the smaller exchange current density that results from the low concentration of electrolyte, and the potential profile in the unsupported electrolyte is also more dispersed. Since the concentration of the supporting anions is much lower, the ionic current in the system is carried mainly by the target anions themselves. The symmetric concentration profile of the target anions (in contrast with the fully-supported case), indicates that both electromigration and diffusion play an important role in the mass transport of the target anions within the electrosorption cell. To elaborate further, near the cathode ($\tilde{y} = 1$), since the total flux of target anions is zero, we have $\widetilde{N}_{y,A}(\tilde{y} = 1) = -\frac{\partial \tilde{c}_A}{\partial \tilde{y}} - z_A \tilde{c}_A \frac{\partial \tilde{\phi}}{\partial \tilde{y}} = 0$, which requires that the diffusive flux (upward) balances the electromigration of the anions (downward) near the electrode surface, as reflected in the negative concentration gradient in the vertical direction near the cathode. Another direct conclusion from the analysis is that $\tilde{c}_A = e^{-z_A \phi}$ near the cathode, which follows the Boltzmann distribution, as is typical in a (nearly) binary electrolyte with a blocking surface.

In some chemical separation processes, the concentrations of the supporting and target anions are comparable. Under these conditions, i.e., $\gamma = 1$, the relative surface coverages of the electrode reflect directly the selectivity of the redox active moieties for the target over the competing species, assuming the equilibrium concentrations do not differ significantly from their inlet stream values. Also, due to the relatively mild concentration ratio between the supporting anions and the target anions, the requirement for the redox active electrode to favor selective adsorption of the target anions can be relaxed greatly and still a similar surface coverage ratio $\Gamma_{\text{ad}}$ as in the fully supported case can be achieved. For the parameter values used in the simulation, as shown in Table 2 and Table 3, the equilibrium adsorption constant of the electrode material chosen in this case equals $K_{\text{ad}} = 12$, which requires as little as $2.5\ V_T = 64$ mV voltage difference to achieve a selectivity of 12 between the target anions versus the supporting anions when in equilibrium. The concentration profile of the supporting anions is more dispersed than that of the target anions due to the weaker interaction with the electrode surface. The electrostatic potential profile also has a centralized region corresponding to the activation of the electrode as it is stabilized by the weakly bound supporting anions, which move faster in the electrosorption cell than the target anions.

These observations for the three cases are consistent with the trends seen in the break-through curves for each of the anions as discussed below.

**Breakthrough Curves**

The separation performance of an electrosorption cell is reflected in the time-dependent effluent concentrations, i.e., the break-through curves. Time is quantified in terms of bed volumes, with 1 bed volume equaling $Gz\ \tilde{t}_d$, and is the total volume of solution introduced to the bed relative to the volume of the bed itself, or equivalently, the actual process time divided by the residence time in the bed. The effluent concentration of the ions can be defined in two ways, either the spatial average, $\bar{\bar{c}}_{i,out} = \int_0^1 c_i dy$ (used in Figure 5) or the mixing cup average, $\bar{c}_{i,out} = \int_0^1 \tilde{u}(\tilde{y}) c_i dy$ (which is used in Figure 6). The two definitions are introduced to distinguish various washing and desorption operation schemes. Here, two washing/stripping streams ($\gamma_{X,\text{wash/des}} = 1, \gamma_{A,\text{wash/des}} = 0$ or $\gamma_{X,\text{wash/des}} = 0, \gamma_{A,\text{wash/des}} = 2$), and operation schemes (batch desorption or stop-flow desorption) are investigated; the simulation results are shown in Figure 5 and Figure 6 respectively. The difference in results obtained using these two definitions during the adsorption step is relatively small, as can be observed in Figure 5 and 6 during the first 8 bed volumes.

For the three inlet concentration ratios of the competing anions, Figure 5 and 6 give the effluent concentrations of the target and supporting anions, as well as the surface coverage of various states during the adsorption, wash and desorption steps respectively. We consider each of the cases in turn.

With a strong supporting electrolyte and dilute target species, $\gamma = 100$, due to stronger affinity of the target anions for the electrode surface, the target species concentration front (red line) lags the inert supporting anions front (blue line). Under this fully-supported case, it takes about 1 bed volume for the supporting anions to exit the cell, but just over 6 bed volumes for the target anions to break-through (to 5% of the inlet concentration). A single pass of the mixture stream achieves a separation factor $\psi_{A/X} = \frac{\theta_{A,eq}}{\theta_{X,eq}} / \frac{c_{A,eq}}{c_{X,eq}} = 1200 = K_{ad}$, which matches the theoretical limit, indicating a very high selectivity towards the target anions that is controlled by the thermodynamic driving force. Figure 5 (b) gives the evolution of surface coverage versus time at the anode for the reduced state $\theta_R$, for the oxidized state that pairs with the supporting anions $\theta_X$, and for the oxidized state that forms a complex with the target anions $\theta_A$. During the adsorption step, two clear time scales are evident. The first time scale occurs within the first bed volume, which indicates the activation of the electrode in the presence of the supporting electrolyte; and the second time scale takes place between the 1 to 8 bed volumes, which corresponds to the formation of a more stable target anion – oxidized state complex. This formation of a more stable complex occurs not only through ion exchange with the less stable supporting anion – oxidized state complex, but also results from a more complete utilization of the electrode via the stronger stabilization of the oxidized electrode by the target anions - the cathodic shift (see SI S2.5 for more discussion) in the equilibrium voltage $\Delta\tilde{\phi}_{eq,A}^{ref} = \Delta\tilde{\phi}_{eq,X}^{ref} - \ln(K_{ad})$ results in additional Faradaic activation of the redox-active moieties, leading to a final surface coverage of the target anions of 86%. By comparison, there is only 57% surface activation after the break-through of the supporting anions in the first bed volume, as the active binding site is stabilized less by the supporting electrolyte than by the target anions.

During the wash step, the applied cell voltage is held constant in order to reduce the loss of the captured target anions to the wash stream. Both wash schemes rely on the reduction of the supporting electrolyte concentration within the cell and a decrease in the surface coverage of the supporting anions to almost zero. In the first operational scheme, the surface coverage of the target anions also decreases due to the shift in the (electro-)chemical equilibrium with the decreased concentration of the target anions in the wash stream, as reflected in Figure 5 (b). Utilization of a wash stream with a higher concentration of the target anions, as proposed in the second operation scheme, not only avoids the undesired loss of the target anions to the wash stream (Figure 6 (a)), but also further activates the electrode surface from 86% to 96%, as shown by the ramping up of surface coverage during the wash step in Figure 6 (b).

During the desorption step, the applied voltage is decreased to return the majority of the redox active surface binding sites to their original uncharged state, and to release the adsorbed anions to the stripping solution. The flow rate is decreased during this step to recover the target anions in a concentrated stream. For the first operation scheme, $\widetilde{U}_{\text{des}} \sim 0$, equivalent to a batch operation. With a pure dilute supporting electrolyte as the stripping stream ($\gamma_{A,\text{wash/des}} = 0, \gamma_{X,\text{wash/des}} = 1$), Figure 5 (b) shows the concentration of the target anions reaches a constant value of 6.2 during the batch desorption step, which is close to $\tilde{c}_{A,\text{des,max}} = 6$ from the theoretical analysis in SI S5.1, achieving a purity level of 86%. The anode surface, meanwhile, is recovered to the original uncharged reduced state and is ready for the next cycle of operation. For the second operation scheme, as elaborated in SI Figure S2 with results shown in Figure 6 (b), the desorption step has a stop-flow pattern, where the flow is stopped for 0.5 bed volume to allow complete desorption in batch, and then flow is resumed for another 1.5 bed volumes. The cell is washed with a higher concentration of pure target anions ($\gamma_{A,\text{wash/des}} = 2, \gamma_{X,\text{wash/des}} = 0$), and the effluent concentration of the target anions reaches as high as 8.5, due to both the higher stripping solution inlet concentration and the further activation of the electrode surface during the wash step. After around 1 bed volume, the released target anions are flushed and collected with a nearly 100% purity, which is very attractive for reducing the cost of downstream processing of the captured target ions.

For the case with unsupported electrolyte, $\gamma = 0.01$, during the adsorption step, it also takes about 6 bed volumes for the target anions to break-through from the ion-adsorption cell. Due to the ultra-low influent concentration of the supporting anions, it takes a very long time for these anions to break through, and thus the surface coverage of the supporting anions is almost unchanged during the adsorption step. The reason for the non-zero surface coverage ($\theta_X \sim 0.15$) of the supporting anions on the electrode is that the system is initialized with a supporting anion concentration at the same level of $c_{A,0}$ (in the micromolar range). In this case, the initial concentration of the supporting electrolyte in the electrosorption cell is higher than its influent concentration, and when a positive voltage is applied, some portion of the supporting anions in the open channel are adsorbed onto the electrode surface. A detailed analysis of, and an analytical solution for the initial surface coverage of the electrode can be found in SI S5.3. After the initial activation by the pre-filled supporting anions within the cell, the current contribution from the Faradaic reaction with the extremely dilute supporting anions is almost zero. Therefore, when the target anion front propagates through the ion adsorption cell, it activates the electrode surface solely due to the stabilizing Faradaic electrosorption process. The redox active moieties are converted from the reduced state to the oxidized state, and there is little contribution from the ion exchange mechanism.

When the concentration of the supporting electrolyte is comparable to that of the target anions, $\gamma = 1$, its concentration is maintained around 0.15 during the first three bed volumes. In comparison with the unsupported electrolyte scenario, the inlet concentration of the supporting anions is just sufficient to activate the electrode surface and maintain a relatively fast propagation velocity within the ion adsorption cell. Between 4 and 7 bed volumes, the concentration of the supporting anions in the effluent stream becomes higher than the inlet concentration, which indicates displacement of the supporting ions by the target species through an ion-exchange process. This is also confirmed by Figure 5 (b), where the surface coverage of the supporting anions drops during this time period. Under the first wash/desorption operation scheme, the surface coverage of the target anions complex dropped slightly due to the decrease in target anion concentrations in the cell. During the (batch) desorption, both the target and the supporting anions are released to the stripping stream, where the final concentrations of the target anion and supporting anions become 6.3 and 1.1, respectively, achieving a product purity of 85%. When the second wash/desorption scheme is used, the pure target anion stripping solution washes out the excess amount of the supporting anions and liberates more surface binding sites for further capture of the target anions, which results in a final surface coverage of the target anions ramping up from 86% to 96% at the end of the wash step. During the desorption step, the captured anions on the electrode surface are released to the cell in a batch process first (stop step) and then pushed out by the stripping solution (flow step), which takes around one bed volume to collect the regenerated anions with a concentration of 8.5 under near 100% purity.

**Adsorption Front Propagation**

In addition to the break-through curves discussed above, which provide an overview of dynamic salt adsorption performance, the detailed modeling framework developed here also enables us to monitor the spatial concentration distributions of the two competing anions within the electrosorption cell. Figure 7 demonstrates the concentrations of the target anions and supporting anions near the anode surface within the flow cell at various times. The uniform spacing between the different curves indicates that the concentration wave front propagates through the channel at a constant velocity. The concentration front of the target anions resembles that of a traveling wave, which can be characterized by a wave velocity and a wave width (the exploitation of these effects to reduce the model to a one-dimensional description is beyond the scope of this paper, and will be addressed in a future report). The constant wave front shape is due to the self-sharpening effect because of the favorable curvature of the adsorption isotherm [66] (the Langmuir isotherm, as derived in SI S2.4), which is essential for reducing the dispersion effects in the electrosorption cell. The traveling wave behavior is described qualitatively here: behind the wave front, due to the fast adsorption kinetics, local equilibrium is established and the equilibrium bulk concentration is close to the inlet anion concentrations; ahead of the wave front, since the surface capacity of the ion adsorption cell is much larger than the bulk concentration of the anions, i.e., $\varepsilon \ll 1$, both the supporting anions and the target anions will be depleted in the channel and the electrosorption process is limited by the convection speed of the ionic species.

For the fully supported case, $\gamma = 100$, it is clear that the adsorption wave fronts of the two competing anions within the first bed volume propagate at different speeds (lines of the same color indicate concentration profiles at given times of $\hat{t} = 0.167$ (blue), 0.5 (red) and 1 (purple) bed volumes), with the target anion front moving much more slowly than that of the supporting anions: this is a direct result of the stronger interaction of the target anions with the adsorption sites of the ion adsorption cell, which is the key consideration in material selection for selective separation applications.

For the scenarios with comparable ($\gamma = 1$) or dilute ($\gamma = 0.01$) supporting electrolyte, since it takes much longer for the supporting anions to break through from the ion adsorption cell, the concentration distributions are plotted over longer time ranges $0.5 \leq \hat{t} \leq 8$, and results at 0.5 (blue), 3 (red) and 5.5 (purple) bed volumes are highlighted. From the spacing of the concentration profiles under $\gamma = 0.01$, it can be seen that the supporting anion wave front propagates at a similar speed as that of the target anions; the shoulder in the front is due to ion-exchange displacement of the supporting anion by the target species When the concentrations of the two competing anions are comparable, $\gamma = 1$, the supporting anions show a mixed Faradaic adsorption and ion-exchange desorption behavior. When the target anions propagate within the ion adsorption cell, they displace the weakly bound supporting anions and result in a stream with a concentration of the supporting anions higher than the inlet concentration. The colored lines clearly show that the supporting anions move faster than the target anions in this case.

In all three scenarios, the concentration front of the target anions propagates in a universal self-similar fashion, which can be visualized by plotting the concentration distribution along both the spatial and time coordinates as shown in Figure S3 Left (SI). It is evident that the concentration profile moves at the characteristic velocity of the adsorption wave. If the spatial and time coordinates are carefully lumped into a single coordinate using the characteristics velocity: $\hat{x}' = \hat{x} - \widetilde{U}_{\text{eff}} \tilde{t}$, then after an initial transient period, the concentration curves all collapse onto a unified curve, as is shown in Figure S3 Right (SI).

**Adsorption Rates**

The competition of ion adsorption between the target and supporting anions on the electrode surface is directly reflected in the different adsorption rates under spatial and time coordinates. Figure 8 demonstrates the adsorption rate of the target anions (a) and the supporting anions (b) along the anode surface at various times.

For $\gamma = 100$, the adsorption rate of the supporting ions has two distinct regimes during the adsorption step. The first regime is $\hat{t} < 1$, when the concentration front of the supporting ions passes rapidly through the channel, as reflected by the positive peaks with large magnitude in Figure 8 (b). The second regime corresponds to $\hat{t} > 1$, where the target anions displace the adsorbed supporting species, leading to a negative supporting ions adsorption rate, as indicated by curves with the same color in the two subplots. Meanwhile, the concentration front of the target anions begins to propagate through the channel, and the adsorption rate profiles adopts a skewed Gaussian shape, indicating the process is limited by mass transport. Ahead of the convection front, the adsorption rate is zero, as the solution entering this region of the channel is depleted of the target anions, while behind the convection front, the local adsorption rate also drops to zero, as the electrode surface is in equilibrium with the target anion feed concentration.

For the unsupported electrolyte, $\gamma = 0.01$, adsorption of the target anions is the dominating mechanism within the ion adsorption cell, which is evident from the magnitude of the ionic fluxes. From $\widetilde{N}_{y,X}(\tilde{y} = 0) = -(\widetilde{R}_{F,X} - \widetilde{R}_{\text{ad}}) \approx 0$, we know that the ion-exchange process is also negligible due to the dilute concentration of the supporting anions. The spacings of the ion adsorption rate curves under various time steps indicate a similar characteristic velocity between the two competing anions.

Under the intermediate inlet concentration ratio, $\gamma = 1$, the magnitudes of ionic fluxes from the two competing ions are also comparable. However, for the supporting anions, the ionic flux is positive (in the spatially averaged sense) for the first three bed volumes. As the target anions propagate within the bed, the ionic flux of the supporting anions becomes negative, indicating an ion-exchange mechanism under the inlet conditions.

For all the three scenarios, the new coordinate: $\hat{x}' = \hat{x} - \widetilde{U}_{\text{eff}} \tilde{t}$ can be used as was done in the adsorption front analysis to lump the adsorption rates of the target anions onto a universal curve, as shown in Figure S4 (SI).

**Surface Coverages of Redox-Active Moieties**

The surface coverages of various states of the redox-active moieties along the length of the ion adsorption bed at different times are shown in Figure 9. For $\gamma = 100$, almost all of the redox centers of the anode are initially in the reduced state. As the mixture stream fills the channel, the surface binding sites are activated and stabilized by the supporting anions, but gradually the more stable surface complexes form with the target anions due to their stronger interactions with the surface binding sites. Depending on the applied voltage and inlet concentration ratio, the ratio of surface coverage of the supporting anions relative to that of the reduced state reaches 1:1 for the fully-supported case, which gives around 50% activation at the end of the first bed volume, as indicated by the grey curve at $\hat{t} = 1$ immediately after the blue curve. After the first bed volume, the target anion concentration front propagates slowly in the cell, as it displaces the pre-adsorbed supporting anions and forms more stable surface complexes on the anode and the reduced state finally drops down to 7%. The surface coverage percentage of the target anions approaches 86% at equilibrium under the applied voltage during the adsorption step. In contrast to the fully supported case, the surface coverage of the supporting anion complex in the unsupported electrolyte, $\gamma = 0.01$, remains constant at a small non-zero value (~0.15), due both to the slow migration speed of the supporting anions under low inlet concentration and to the adsorption on the electrode surface of the initial amount of supporting anions within in the cell when the process starts up (SI S5.3). Therefore, the main adsorption mechanism is the activation of the electrode surface due to the target anions themselves.

For the intermediate case, with $\gamma = 1$, it takes longer (around 3 bed volumes) for the supporting anions to reach 50% surface coverage, which eventually drops to 7% due to ion-exchange replacement by the target anions. The final surface coverage of the target anions reaches 86%, similar to the value in the first case.

**Current Density and Adsorption Rate**

The average current density across the entire electrode surface is an easily accessible, experimentally measurable signal of the electrosorption process. Figure 10 shows the surface averaged current response $\bar{I} = \int_0^{\hat{L}} \tilde{\imath}/\hat{L}$ of the ion adsorption cell (blue) and compares it with the adsorption rate of the target anions $\bar{N}_A = \int_0^{\hat{L}} \widetilde{N}_A / \hat{L}$ (red). The popular belief that the current response indicates the ion adsorption rate is not true in the case of multiple ions competing for the binding sites on the same electrode surface.

It is evident from the figure that, for $\gamma = 100$, the current response has two stages, the first with the larger magnitude corresponds to the activation stabilization of the electrode by the supporting electrolyte; the second stage reflects the further activation of the redox active moieties on the electrode surface due to their additional activation and stabilization by the target anions. However, for the target ions, due to their slower effective wave velocity, the surface averaged adsorption rate demonstrates a more sluggish dynamics.

As discussed earlier in relation to Figure 8, in the unsupported case, $\gamma = 0.01$, the flux contribution from the supporting anions to the anode is small, and the ion adsorption rate from the Faradaic reaction stabilized by the supporting anions is negligible, so that the total current density reflects primarily the ion adsorption rate of the target anions, and the surface averaged current density and the ionic flux of the target anions towards the anode match well with each other after a short transient period.

The current response of the third scenario, $\gamma = 1$, looks similar to that for the strong supporting electrolyte, where the current signal exhibits two regimes, the first corresponding to the activation of the anode by the supporting anions and the second to the activation by the target anions. However, the lower concentration of the supporting anions (compared with the first case) results in a slower propagation speed within the cell and a longer break-through time (around three bed volumes in Figure 5), which agrees well with the length of the first current plateau. The adsorption rate of the target anions has a relatively constant magnitude, which corresponds to the rate of the second-stage Faradaic activation as well as the ion exchange process, which is limited by mass transport of the target anions.

**Current Density Distribution**

Earlier discussions have demonstrated that the space and time dependent concentration distributions of the ionic and surface species within the electrosportion cell are essential in understanding the interplay between bulk mass transport and competitive surface adsorption on the electrode. However, it is very challenging to measure the spatially and temporally dependent concentration profiles experimentally. Spatial current densities are relatively easier accessible signals if spatially segmented electrodes are utilized [67]; such measurements would add another dimension of information and indicate the mass transport progress (especially for the target anions) within the ion adsorption cell, which would be difficult to obtain otherwise. To visualize this approach, the anode was segmented to consist of six electrodes distributed over the channel length. The average current response of individual segments is plotted versus time in Figure 11. The current responses from the six channels for $\gamma = 100$ show similar features to those noted above, with two distinct regimes that correspond to the filling of the cell with the supporting electrolyte and subsequently to the slower propagation of the target anion adsorption front in the cell. The first peak is sharp and has high intensity, which corresponds to the concentration wave front of the supporting anions, while the second peak is much broader and indicates the concentration front of the target anions; since the electrosorption rate of the supporting anions during the second regime is close to zero, i.e. $\tilde{R}_{F,X} = 0$, the total current is a direct indicator of the electrosorption rate of the target anions $\tilde{I} = \tilde{R}_{\text{F},A}$. This reconfirms our previous statement about the distinction between the current response and the true adsorption rate of target anions. The time spacing between the adjacent current peaks in the second regime corresponds well with the effective wave velocity of the target anions, which again indicates that it is the secondary current response that represents the concentration wave front of the target anions.

For the unsupported electrolyte, $\gamma = 0.01$, the current response from the segmented electrodes also shows distinct peaks from each current collector and indicates propagation of the concentration wave front of the target anions (primarily) in the ion adsorption cell. In contrast to the fully supported case, the sharp peaks with large magnitude in the first bed volume no longer exist due to the lack of sufficient supporting anions to activate the electrode surface before the target anions fill the channel. The single time scale of around 6 bed volumes corresponds to the activation of the electrode surface by the target anions themselves. The locations of the current peaks can also be used to calculate the effective wave velocity, which agrees well with the theoretical prediction.

The current response from the segmented electrodes for the last scenario, $\gamma = 1$, also shows two peaks, similar to the first case. The second peak is not evident until the third electrode segment, because the two competing anions move at a similar speed, which is not the case in the fully supported scenario, in a longer time or longer bed length is taken for the two concentration fronts to be separated.

**Conclusion**

We have formulated and analyzed a general mathematical model for electrosorption systems with Faradaically modulated redox active electrodes, which can be used to understand and optimize the selective removal of target ions from a mixture of competing species in wastewater remediation and chemical separation applications. A two-dimensional transient model is formulated from first principles with no tuning parameters. The model incorporates thermodynamic properties, electron transfer kinetics derived from non-equilibrium thermodynamics, and mass transport limitations (from diffusion, convection and electromigration) during the ion adsorption process. The dimensionless equations of the model reveal the importance of several dimensionless groups, such as $Gz, \varepsilon, \gamma, Da, \nu_A, \Gamma_{\text{ad},A}$ and $\Gamma_{\text{ad},X}$, which govern five physically relevant time scales corresponding to diffusion, convection, reaction, surface saturation and adsorption front propagation, and completely describe the behavior and performance of the competitive ion adsorption processes.

For selective adsorption applications, we are particularly interested in the operating regime in which the propagation time scale of the target anions within the ion adsorption cell is the slowest process (limiting step). This allows us to focus on a smaller design space with the operation parameters. Through analytical derivations and numerical simulations, we demonstrate that for sufficiently thin channels, when $Gz(Da + \nu_A e^{(1-\alpha)\Delta\tilde{\phi}_{\text{ad}}}) \gg 1$, the convection limited regime is achieved for the target anions, and the surface transport in the electrosorption cell exhibits a moving front kinetics similar to the transport effects in chromatographic columns or packed bed adsorption. The full model has no restriction on the range of parameter values, of course, and is still valid for other operation regimes of interest, such as $Gz \ll 1$, which is typical in measuring the reaction rate in SPR measurement of biological systems or in energy conversion applications, such as flow batteries.

Within the refined design space, such that $Gz \gg 1$, $\varepsilon \ll 1$, $\Gamma_{ad,A} \gg 1$, $\Gamma_{ad,X} \approx 1$ and $\Gamma_{des,A} \ll 1$ etc., the adsorption/desorption processes are simulated for three different cases to include all scenarios of practical interest for selective adsorption applications, with inlet concentration ratio between the supporting anions and the target anions ranging over four orders of magnitude (0.01 to 1 to 100). A key design parameter of the redox active materials is the equilibrium adsorption constant for the ion exchange reaction, $K_{ad}$. This parameter links the equilibrium voltage of the two Faradaic reactions and governs the selectivity of the process by the expression: $S_{A/X} \approx K_{ad} = e^{-\Delta \tilde{E}_{ad}} = e^{-(\Delta \tilde{\phi}^\Theta_{eq,A} - \Delta \tilde{\phi}^\Theta_{eq,X})}$. It is also derived analytically that, in order to keep a fixed surface coverage ratio at equilibrium after the adsorption step, $K_{ad}$ should scale linearly with the inlet concentration ratio, i.e., $K_{ad} \approx \frac{\Gamma_{ad,A}}{\Gamma_{ad,X}} \gamma$ which allows a relaxation on the material selection under less challenging separation conditions. Under all scenarios, the convection limited regime is observed for the target anions, and the favorable curvature of the adsorption isotherm results in the self-sharpening concentration wave front.

The simulation results demonstrate that two distinct operation regimes occur during the adsorption step. The first corresponds to the activation of the electrode surface by the weakly bound target anions that move faster in the cell, and this process takes about 1 to 3 bed volumes depending on the inlet concentration ratios between the two competing anions; the second regime derives from the further activation and stabilization of the redox active moieties on the electrode surface by the target anions due to the thermodynamic driving force, and this process continues until the target anions break through from the electrosorption cell. Depending on the initial concentration ratios, the ion-exchange effect from the competing anions can play an important ($\gamma = 100$ or $1$) or negligible role ($\gamma = 0.01$).

We have also proposed a three-step stop-flow operation, where an intermediate wash step with pure target anions is used to wash away the excess amount of the supporting anions, and further activate and replace the electrode binding sites with the target anions; the stop-flow desorption step allows the regeneration of the target anions under high concentration. In comparison with using a dilute supporting electrolyte as the wash/stripping solution, the proposed strategy enables an increase in the surface coverage of the target anions from 86% to 96%, and the purity of the regenerated stream increases from 86% to almost 100%; both effects are beneficial for downstream processing or storage.

The mathematical framework developed here provides general engineering criteria for the design and operation of continuous adsorption units based on diffusion, convection, electromigration and reaction of multiple competing ions under an electric field, and has applications not limited to electrochemical water treatment, but also to energy storage, microfluidic reactor design and bio-sensing, etc. However, the full two-dimensional time dependent model is computationally expensive to solve, taking about 40 mins to complete a full cycle of operation (with adsorption, wash and desorption steps) under a given set of parameters. A reduced order model that captures the key physics of the electrosorption system could save on computational efforts, and would be more readily used to extract model parameters from experimental data, and facilitate future optimization and scale-up of the electrosorption processes, which is the focus of a companion paper [68].

**Figures**

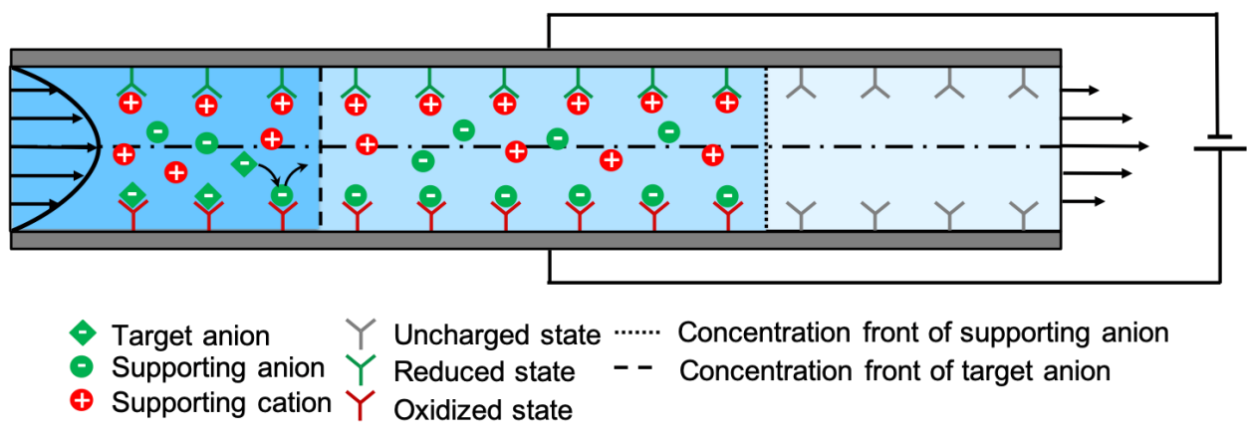

- ◆ Target anion
- ● Supporting anion
- ⊕ Supporting cation
- Y Uncharged state
- Y Reduced state
- Y Oxidized state
- ······ Concentration front of supporting anion
- -- Concentration front of target anion

Figure 1. Schematic of the Faradaically Modulated Selective Adsorption process. Anode locates at the bottom and cathode at the top and an electric voltage is applied across the cell. Both electrodes are modified with redox active materials and are activated under the electric field. The long and short dashed lines indicate the concentration wave fronts of the target and supporting anions respectively. Due to the favorable binding between the target anion and the electrode surface, it displaces the supporting anion and propagates at a much slower velocity in the cell.

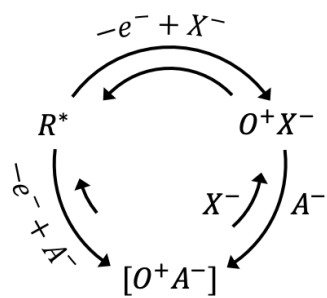

Figure 2. Reaction network of the competitive electrosorption at the anode. $R^*$ denotes the uncharged reduced state, $O^+X^-$ and $[O^+A^-]$ denotes the surface complex of the oxidized state electrostatically paired with the supporting anions and the oxidized state chemically bound with the target anions. The length of the arrowed arcs indicates the magnitude of forward/backward reaction rate constants.

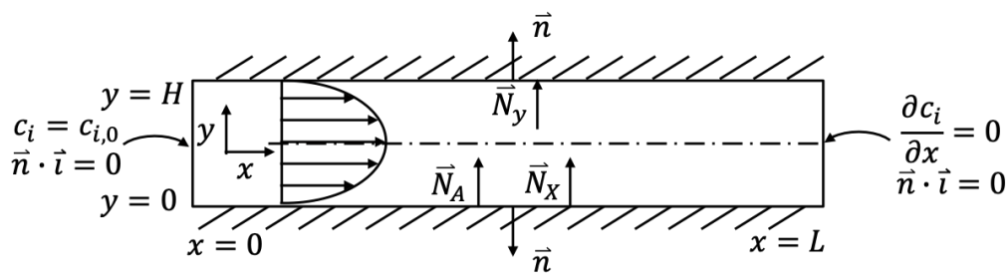

Figure 3. Schematic of the two-dimensional cell with boundary conditions. The height and length of the ion adsorption cell is $H$ and $L$ respectively. The normal direction points outwards the boundary. The dot-dashed line denotes the geometric symmetry line.

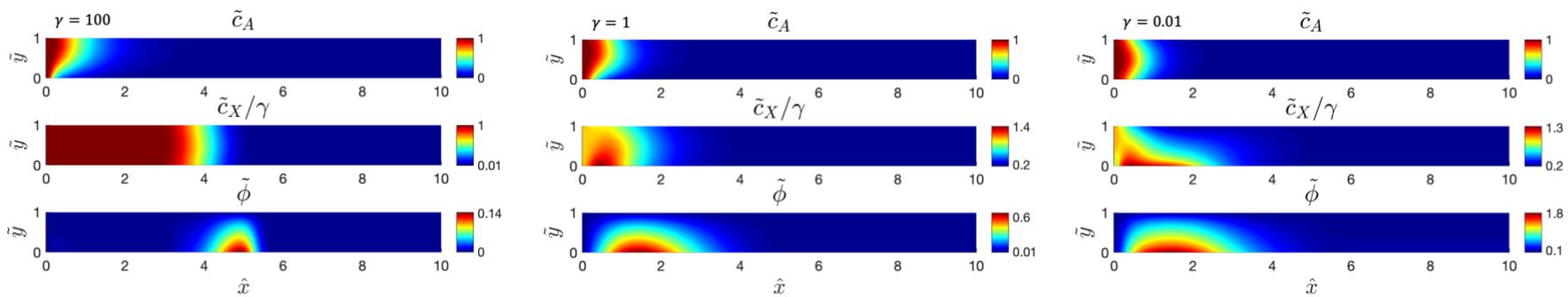

Figure 4. Concentration of target anion, supporting anion and electrostatic potential at $\hat{t} = 0.5$. (The concentration of the supporting anions is normalized by the inlet concentration ratio $\gamma$.)

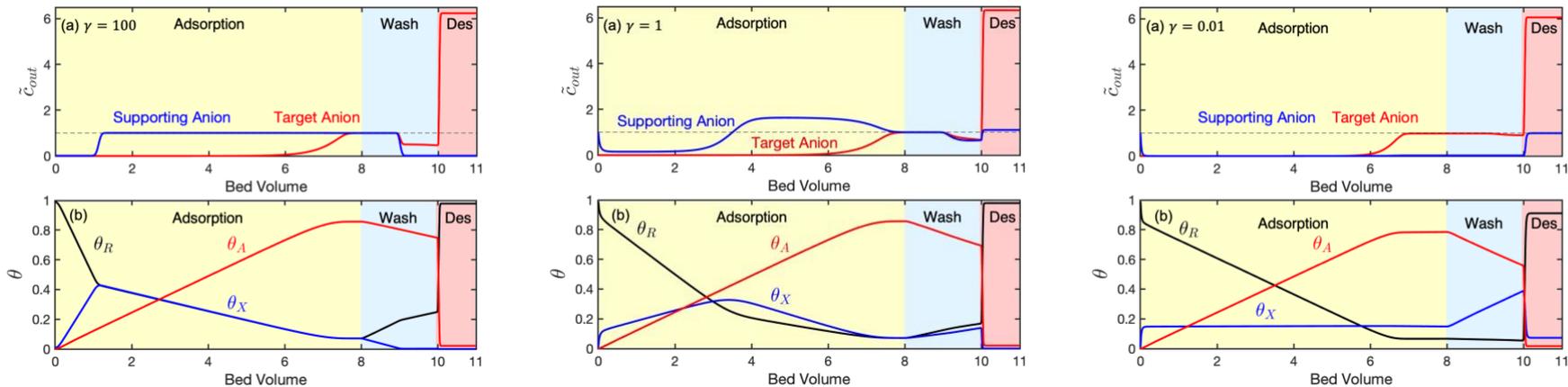

Figure 5. (a) Effluent concentration of target and supporting anions (normalized by $\gamma$ when $\gamma = 100$) vs. bed volume (time). (b) Surface coverage of various states vs. bed volume (time). Time range for each step: adsorption 0-8, wash 8-10, desorption (in batch) 10-11 bed volumes.

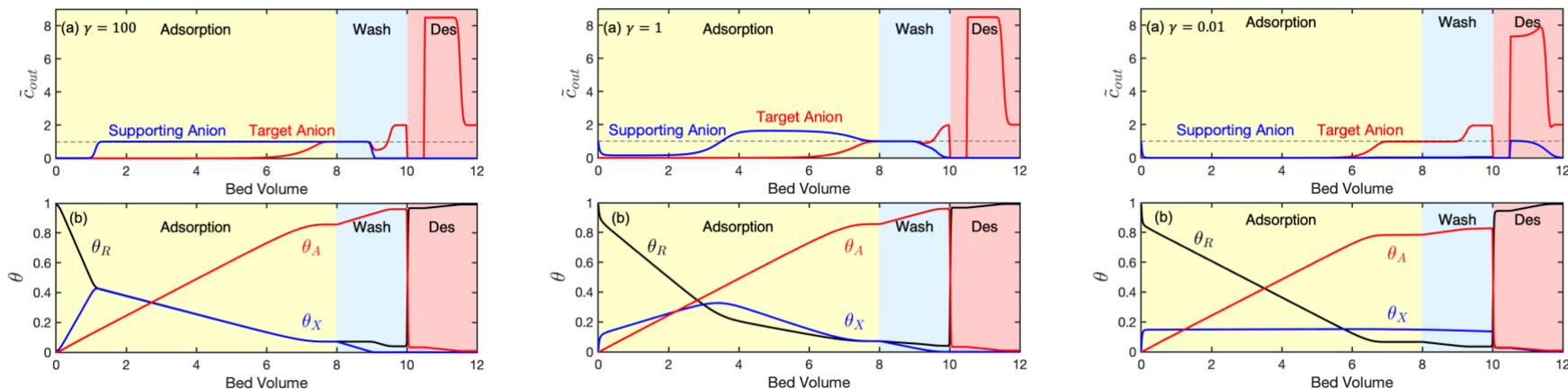

Figure 6. (a) Effluent concentration of target and supporting anions (normalized by $\gamma$ when $\gamma = 100$) vs. bed volume (time). (b) Surface coverage of various states vs. bed volume (time). Time range for each step: adsorption 0-8, wash 8-10, desorption (stop 0.5, flow 1.5) 10-12 bed volume.

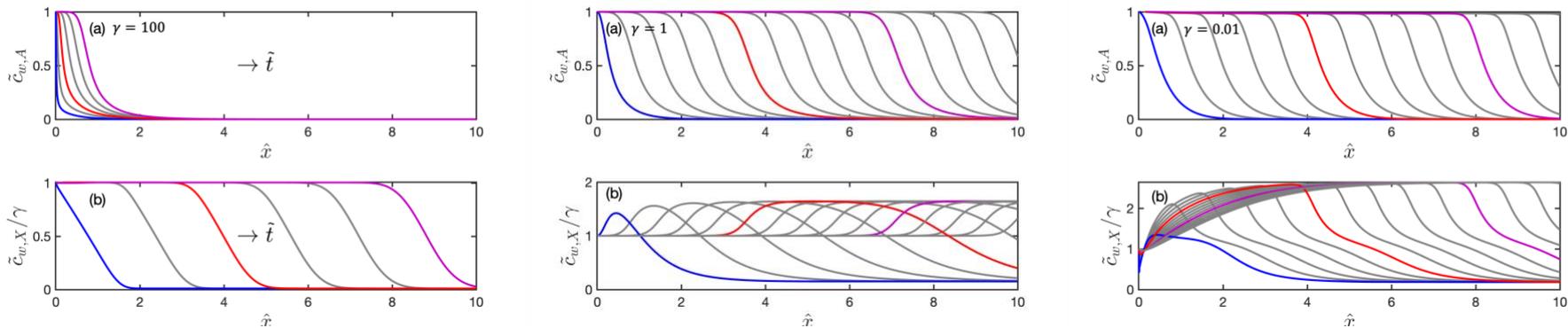

Figure 7. Concentration of (a) target anions, (b) supporting anions (normalized by $\gamma$) along the anion bed length under different time. For $\gamma = 100$, the colored lines indicate $\hat{t} = 0.167$ (blue), 0.5 (red) and 1 (purple); for $\gamma = 1$ or 0.01, they correspond to $\hat{t} = 0.5$ (blue), 3 (red) and 5.5 (purple).

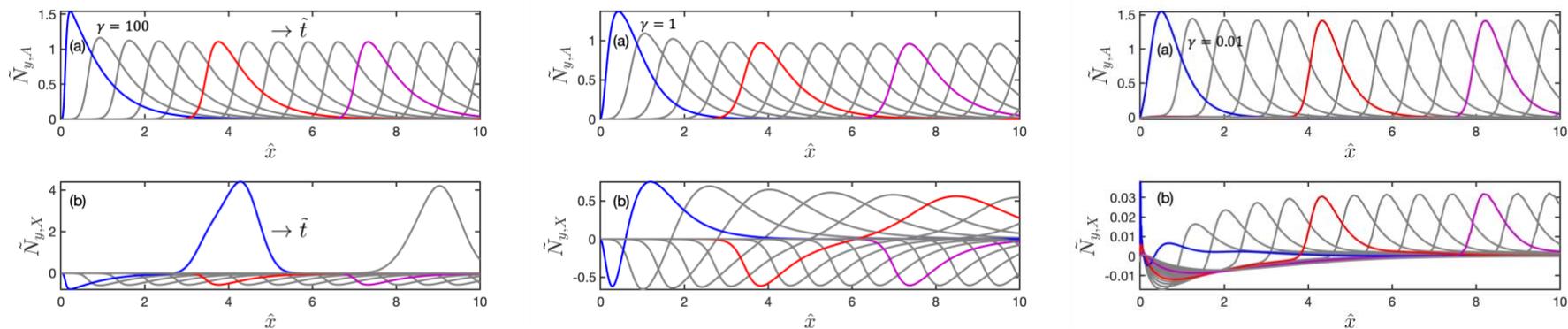

Figure 8. Adsorption rate of (a) target anions (b) supporting anions at the anode along the bed length under different time, $\hat{t} = 0.5, 1, \ldots, 8$. The colored line indicated the adsorption rate at $\hat{t} = 0.5$ (blue), 3 (red) and 5.5 (purple).

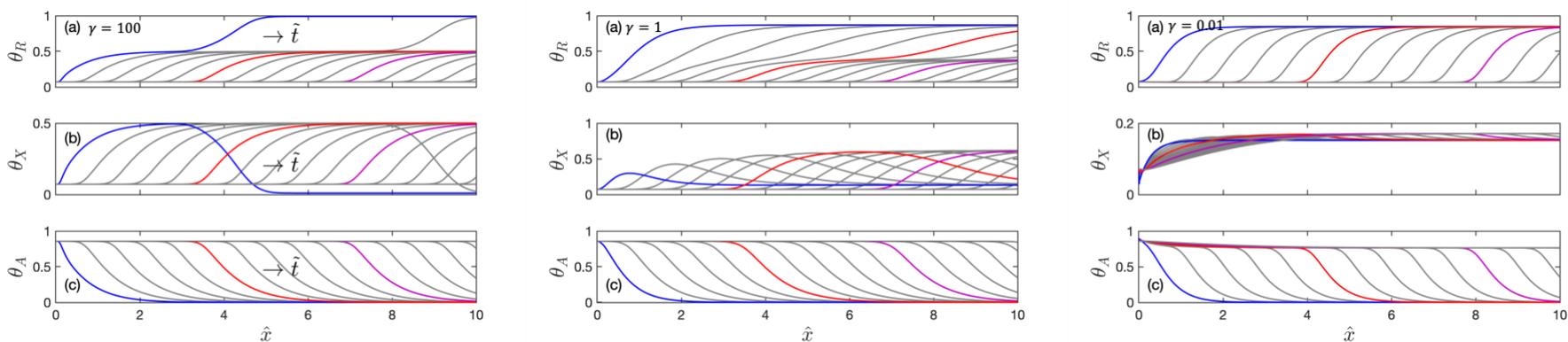

Figure 9. Surface coverage at anode versus bed length at various time (a) reduced state, (b) oxidized state - supporting anions complex, (c) oxidized state – target anion complex. $\hat{t} = 0.5, 1, \ldots, 8$. The colored lines indicate the surface coverage at $\hat{t} = 0.5$ (blue), 3 (red) and 5.5 (purple).

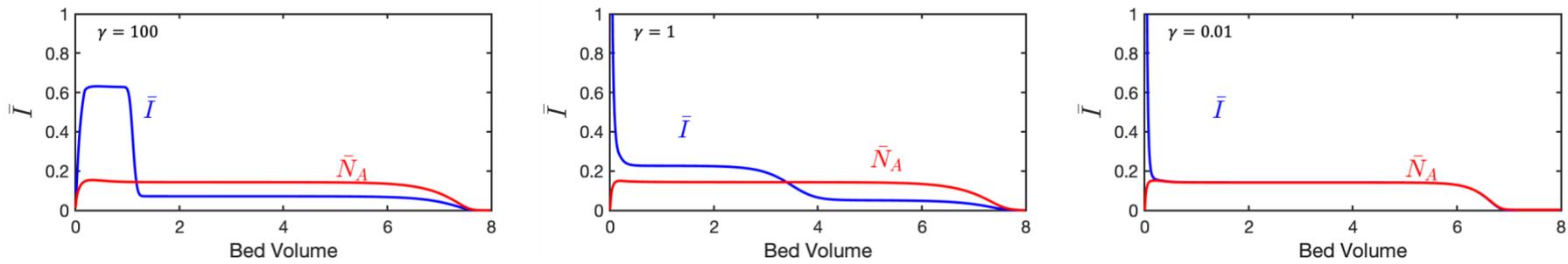
Figure 10. Surface averaged current density and adsorption rate of the target anions vs. bed volume (time) during the adsorption step.

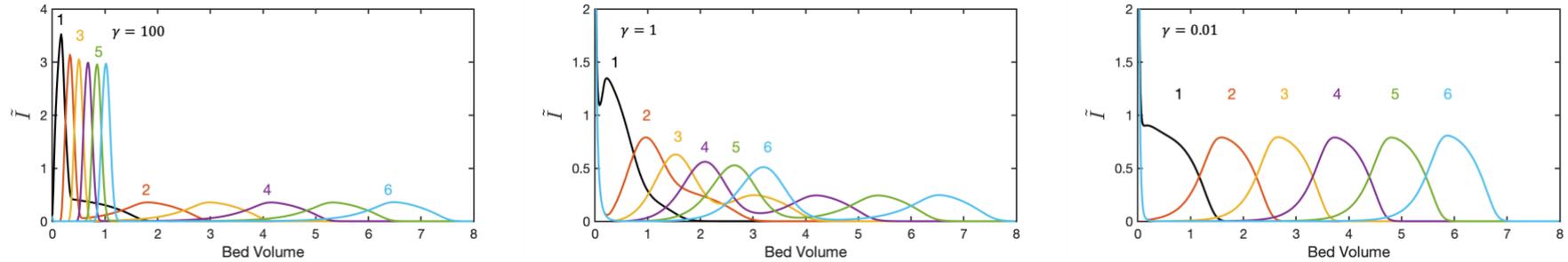
Figure 11. Multichannel current response during the adsorption step with six segments of electrodes, each with a length of one sixth of the total bed length. The current response is the surfaced averaged current for each individual electrode segment

**Tables**

Table 1. Boundary conditions for the three-step stop-flow operation

|  | Adsorption | Wash | Desorption | |
|---|---|---|---|---|
|  |  |  | Stop | Flow |
| $\tilde{u}(\tilde{y})(\hat{x},\tilde{y},\tilde{t})/6\tilde{y}(1-\tilde{y})$ | 1 | 1 | $\tilde{U}_{des}$ | 1 |
| $\tilde{c}_A(\hat{x}=0,\tilde{y},\tilde{t})$ | 1 | $\gamma_{A,wash}$ | $\gamma_{A,des}$ | |
| $\tilde{c}_X(\hat{x}=0,\tilde{y},\tilde{t})$ | $\gamma$ | $\gamma_{X,wash}$ | $\gamma_{X,des}$ | |
| $\tilde{V}_{cell}(\hat{x},\tilde{y}=0,\tilde{t})-\Delta\tilde{\phi}_{eq,A}^{ref}$ | $\Delta\tilde{\phi}_{ad}$ | $\Delta\tilde{\phi}_{ad}$ | $\Delta\tilde{\phi}_{des}$ | |

Table 2. Parameter values in simulation of Electrochemically mediated selective adsorption

| Parameter | Value | Definition |
|---|---|---|
| $D_A$ | $10^{-9}$ [m$^2$s$^{-1}$] | Diffusivity of target anion A |
| $D_X$ | $10^{-9}$ [m$^2$s$^{-1}$] | Diffusivity of supporting anion X |
| $D_Y$ | $10^{-9}$ [m$^2$s$^{-1}$] | Diffusivity of supporting cation Y |
| $z_A$ | -1 | Charge of anion A |
| $z_X$ | -1 | Charge of supporting anion X |
| $z_Y$ | 1 | Charge of supporting cation Y |
| $U$ | $2\times10^{-3}$ [ms$^{-1}$] | Flow rate |
| $H$ | 100 [μm] | Thickness of channel |
| $L$ | 0.12 [m] | Length of channel |
| $c_{s,0}$ | $7\times10^{-5}$ [mol m$^{-2}$] | Surface site density |
| $k_f$ | 1.2 [m$^3$ mol$^{-1}$ s$^{-1}$] | Forward adsorption rate |
| $k_b$ | $k_f/K_{ad}$ [m$^3$ mol$^{-1}$ s$^{-1}$] | Backward adsorption rate |
| $c_{A,0}$ | 0.1[mM] | Inlet concentration of A |
| $c_{X,0}$ | $\gamma c_{A,0}$[mM] | Inlet concentration of X |
| $c_{X,init}$ | 0.1[mM] | Initial concentration of X |
| $c_{ref}$ | 0.1[mM] | Reference concentration for normalization |
| $c^\Theta$ | 1[M] | Standard state concentration |
| $k_A^0$ | 0.518 [ s$^{-1}$] | Faradaic reaction rate constant for A |
| $k_X^0$ | 0.518 [ s$^{-1}$] | Faradaic reaction rate constant for X |
| $\alpha$ | 0.5 | Charge transfer coefficient |
| $\Delta\phi_{eq,X}^\Theta$ | 0.020 [V] | (Standard state) equilibrium potential of X vs cathode |
| $\Delta\phi_{eq,X}^{ref}$ | 0.257 [V] | (Reference state) equilibrium potential of X vs cathode |
| $\Delta\phi_{eq,A}^{ref}$ | $\Delta\phi_{eq,X}^{ref}-\ln(K_{ad})$ [V] | (Reference state) equilibrium potential of A vs cathode |
| $\Delta\phi_{eq,Y}^{ref}$ | 0 [V] | (Reference state) cathode equilibrium potential shifted to 0 |

Table 3. Representative values of dimensionless groups in the simulation

| Dimensionless group | Value | Definition |
|---|---|---|
| $\beta$ | 1200 | Aspect ratio (length vs. height) of the channel |
| $Pe$ | 200 | Peclet number |
| $Da$ | 8.4 | Damkohler number for ion exchange reaction |

| Symbol | Value | Description |
|---|---|---|
| $\nu_A$ | 36.3 | Dimensionless rate for Faradaic reaction with A |
| $\nu_X$ | 36.3 | Dimensionless rate for Faradaic reaction with X |
| $\varepsilon$ | 0.143 | Bulk to surface capacity |
| $Gz$ | 6 | Graetz number |
| $K_{\text{ad}}$ | $\dfrac{\Gamma_{\text{ad},A}}{\Gamma_{\text{ad},X}}\gamma$ | Adsorption equilibrium constant |
| $\gamma$ | $10^{-2}, 1, 10^2$ | Inlet concentration of X vs $c_{A,0}$ during adsorption |
| $\gamma_{A,\text{wash}}$ | 2 | Inlet concentration of A vs $c_{A,0}$ during wash |
| $\gamma_{X,\text{wash}}$ | 0 | Inlet concentration of X vs $c_{A,0}$ during wash |
| $\gamma_{A,\text{des}}$ | 2 | Inlet concentration of A vs $c_{A,0}$ during desorption |
| $\gamma_{X,\text{des}}$ | 0 | Inlet concentration of X vs $c_{A,0}$ during desorption |
| $\widetilde{U}_{\text{des}}$ | 0 | Flow rate ratio of desorption vs adsorption |
| $\tilde{c}_{A,\text{eq,ad}}$ | 1 | Equilibrium concentration of A during adsorption |
| $\Gamma_{\text{ad},A}$ | 12 | Equilibrium ratio between $\theta_A$ vs $\theta_R^*$ for adsorption |
| $\Gamma_{\text{ad},X}$ | 1 | Equilibrium ratio between $\theta_X$ vs $\theta_R^*$ for adsorption |
| $\Gamma_{\text{des},A}$ | 0.032 | Equilibrium ratio between $\theta_A$ vs $\theta_R^*$ for desorption |
| $\Delta\tilde{\phi}_{\text{ad}}$ | $\ln\left(\dfrac{\Gamma_{\text{ad},A}}{\tilde{c}_{A,\text{eq,ad}}}\dfrac{\tilde{c}_{X,\text{eq,ad}}}{\gamma}\right)$ | Anode potential during adsorption ref. to anion A |
| $\Delta\tilde{\phi}_{\text{des}}$ | $\ln\left(\dfrac{\Gamma_{\text{des},A}}{\tilde{c}_{A,\text{eq,des}}}\right)$ | Anode potential during desorption ref. to anion A |
| $\tilde{V}_{\text{cell,ad}}$ | $\Delta\tilde{\phi}_{\text{eq},X}^{\text{ref}} + \ln(\Gamma_{\text{ad},X}/\gamma)$ | Cell voltage during adsorption |
| $\tilde{V}_{\text{cell,des}}$ | $\tilde{V}_{\text{cell,ad}} - 8$ | Cell voltage during desorption |
| $\tilde{c}_{X,\text{init}}$ | 1 | Initial supporting electrolyte concentration |
| $\Gamma_{\text{init},X}$ | 0.01 | Initial ratio between $\theta_X$ vs $\theta_R^*$ |
| $\theta_{X,\text{init}}$ | 0.0099 | Initial surface coverage of supporting electrolyte |
| $a$ | 1.667 | Scaling factor of cell length to keep aspect ratio 10:1 |